\documentclass[11pt]{article}

\usepackage{latexsym,cite,amsmath,epsfig,amsfonts}

\DeclareFontFamily{OT1}{rsfs10}{}
\DeclareFontShape{OT1}{rsfs10}{m}{n}{ <-> rsfs10 }{}
\DeclareMathAlphabet{\mathscript}{OT1}{rsfs10}{m}{n}

\numberwithin{equation}{section}

\newcommand{\be}{\begin{equation}}
\newcommand{\ee}{\end{equation}}
\newcommand{\nn}{\nonumber}
\newcommand{\bea}{\begin{eqnarray}}
\newcommand{\eea}{\end{eqnarray}}

\newcommand{\ns}{\normalsize}
\newcommand{\pt}{\partial}

\def\a{\alpha}
\def\b{\beta}
\def\g{\gamma}

\def\d{\delta}
\def\e{\epsilon}

\def\f{\phi}

\def\l{\lambda}
\def\m{\mu}
\def\n{\nu}

\def\p{\pi}

\def\r{\rho}
\def\s{\sigma}

\def\t{\tau}

\def\x{\xi}

\begin{document}

\begin{titlepage}

\vspace{-2cm}

\title{
\hfill{\ns OUTP-01-32P}\\
\hfill{\ns SUSX-TH/01-028\\}
\hfill{\ns hep-th/0106190\\[.5cm]}
{\LARGE Brane--World Inflation and the Transition to Standard Cosmology}}
\author{{\ns\large Andr\'e Lukas$^1$ and David Skinner$^2$} \\[0.8em]
      {\ns $^1$Centre for Theoretical Physics,
     University of Sussex}\\[-0.2em]
      {\ns Falmer, Brighton BN1 9QJ, United Kingdom} \\[-0.2em]
      {\it\ns A.Lukas@sussex.ac.uk}\\[-0.2em]
	\\
      {\ns $^2$Department of Physics, Theoretical Physics, 
     University of Oxford}\\[-0.2em]
      {\ns 1 Keble Road, Oxford OX1 3NP, United Kingdom} \\[-0.2em]
      {\it\ns skinner@thphys.ox.ac.uk}}
   
\date{}

\maketitle

\begin{abstract}
In the context of a five--dimensional brane--world model motivated from
heterotic M--theory, we develop a framework for potential--driven
brane--world inflation. Specifically this involves a classification of
the various background solutions of $(A)dS_5$ type, an analysis of
five--dimensional slow--roll conditions and a study of how a transition
to the flat vacuum state can be realized. It is shown that solutions
with bulk potential and both bane potentials positive exist but are
always non--separating and have a non--static orbifold.  It turns out
that, for this class of backgrounds, a transition to the flat vacuum
state during inflation is effectively prevented by the rapidly
expanding orbifold. We demonstrate that such a transition can be
realized for solutions where one boundary potential is negative.
For this case, we present two concrete inflationary models which
exhibit the transition explicitly.
\end{abstract}

\thispagestyle{empty}

\end{titlepage}


\section{Introduction}
\label{intro}

Over the past couple of years there has been much interest in the
cosmological consequences of models in which the standard model of
particle physics is located on a brane embedded in a
higher--dimensional bulk. So far, the most popular of these models
have been ones in which the universe has topology
$\mathcal{M}_{4}\times\mathcal{S}^1/\mathbb{Z}_2$ motivated by
Ho\v{r}ava--Witten
theory~\cite{Horava:1996qa,Horava:1996ma,Witten:1996mz} compactified
on a Calabi--Yau
3--fold~\cite{Lukas:1999yy,Lukas:1999tt,Lukas:1999hk}. To achieve
gauge--unification at $\sim 10^{16}$ GeV and a Planck scale of $\sim
10^{19}$ GeV, this theory requires that the orbifold radius is roughly
two orders of magnitude larger than the length scale of the
Calabi--Yau~\cite{Witten:1996mz}. Furthermore, the orbifold is
currently at an energy scale which may have been accessible during
inflation. If the scale of extra dimensions is indeed $\sim
10^{14}$ GeV or higher, ground--based accelerators seem to have little
prospect of directly observing their effects in the forseeable
future. Therefore there is great interest in looking for cosmological
signatures that are likely to arise from the different evolution in
the five-- and four--dimensional regimes.

For the reason mentioned above, inflation seemed a natural place to
begin the investigation. The first steps were taken in
Ref.~\cite{binf} where explicit five--dimensional background solutions
were derived for the case of an empty bulk, with the branes each
supporting scalar potentials. More general solutions including bulk
potentials were found in {\it e.g.}
Refs.~\cite{Nihei:1999}-\cite{Ochiai:2000}. These solutions couple to
constant bulk and brane potentials and, therefore, relate to realistic
brane--world inflation in much the same way four--dimensional de
Sitter space relates to four-dimensional slow--roll inflation.

The main goal of the present paper is to develop a more realistic framework
for brane--world inflation. Our procedure to do this will be analogous
to the one employed in the four--dimensional context, where three major
steps are involved. It may be helpful in the following to briefly recall
what these steps are in the familiar four--dimensional setting.
First, one needs to find a cosmological solution which couples
to a constant positive potential. In four dimensions, the only possibility
is $dS_4$, of course. Secondly, one needs to find approximate solutions
``near'' $dS_4$ with a slowly varying scalar field potential, in short,
one needs to analyze the slow--roll approximation. Thirdly, one needs
to find a way to end inflation and the theory needs to relax into its
vacuum state corresponding to a vanishing scalar potential. While
this is quite familiar and straightforward in four dimensions, the
analogous steps in five dimensions are far less trivial, as we will see.

Our analysis will be performed in the context of the five--dimensional
brane--world model first introduced in Ref.~\cite{binf}. As mentioned,
this model is based on the orbifold $S^1/Z_2$ and is motivated by
Calabi--Yau compactifications of heterotic
M--theory~\cite{Lukas:1999yy,Lukas:1999tt,Lukas:1999hk}.  In the
simple version to be used in this paper, the model has a single bulk
scalar field $\f$ with bulk potential $V(\f )$ and one scalar $\f_i$
with potential $V_i(\f, \f_i)$ on each of the two boundaries labeled
by $i=1,2$. Note that we generally take the boundary potential $V_i$
to depend on the bulk scalar $\f$, as well as on the boundary scalar
$\f_i$.

It is important to discuss the various vacuum states with
four--dimensional Poincar\'e symmetry and the associated
four--dimensional effective theories of this model. It is those
effective four--dimensional theories that we expect to describe the
evolution of the late universe and which are vital to extract
concrete predictions from brane--world models. For simplicity we
restrict ourselves to vacuum states which couple to constant
potentials $V$, $V_i$. Then there are two types of vacua which arise
for specific values of the potentials, namely five--dimensional
Minkowski space for $V=V_1=V_2=0$ and the warped vacuum due to Randall
and Sundrum~\cite{RS1} for $V_1 = - V_2 = \pm\sqrt{12|V|}$. In this
paper, for simplicity, we will be mostly concerned with the flat
Minkowski vacuum. This also seems appropriate in the M--theory context
since the flat vacuum can be viewed as an approximation to the
domain wall vacuum of heterotic M--theory~\cite{Lukas:1999tt}. However, most
of the methods and ideas presented in this paper should analogously
apply to warped vacuum states.

Let us now return to the three steps towards a realistic framework for
inflation mentioned above and discuss them separately in the context
of our brane--world model. First we need to find cosmological
solutions which couple to constant potentials $V$, $V_i$. Clearly
there is a variety of possibilities and, hence, the situation is far
more complicated than in four dimensions. In this paper, we will focus
on solutions which are part of $dS_5$ or $AdS_5$, for simplicity.
Again, we expect that our ideas can analogously be applied to the
Schwarzschild-$(A)dS_5$ solutions found in Ref.~\cite{BCG,Ida} and it
would be interesting to carry this out explicitly.  We will
systematically classify the solutions of $(A)dS_5$ type particularly
in relation to a number of properties essential for our subsequent
discussion, such as the existence of solutions with static orbifold
and the allowed ranges for the potentials $V$, $V_i$.

\vspace{0.4cm}

Secondly, we will be implementing slow--roll of the various scalar fields
for the most interesting of those solutions. While there are some discussions
of brane--world slow--roll in the literature~\cite{MWBH,CLL} they generally
focus on slow-roll on a single brane. Here, we will consider the
the full five-dimensional background where we allow for slow--roll of
both boundary scalars as well as the bulk scalar. Particularly the latter,
being a five--dimensional field, complicates matters significantly and
makes the analysis much more involved than it is in four dimensions. 

\vspace{0.4cm}

Finally, we need to understand how to end inflation and approach the
(flat) vacuum state in order to make contact with the four--dimensional
effective theory that governs the late evolution of the universe. In
this context, it is perhaps worth pointing out that the background
solutions under discussion do not have an effective four-dimensional
description for most values of the potentials $V$,$ V_i$. That is to
say, generically, they cannot be understood as solutions of an
effective four--dimensional theory which involves a finite number of
(zero mode) fields. The reason for this is of course the existence of
the brane potentials $V_i$ which constitute sources localized in the
orbifold and which excite Kaluza--Klein modes in a non--thermal,
coherent way. Clearly, unlike thermal excitations, these coherent
modes are not diluted away by inflation. It is only if we go to a
region of solution space where the potentials $V_i$ (and, hence, the
coherent Kaluza--Klein excitations) are sufficiently small (in a sense
to be made precise later) that we can hope for an effective
four-dimensional description. All this is in contrast to inflation in
a more traditional Kaluza-Klein setting without branes. There, the
Kaluza-Klein modes can always be set to zero consistently and
the dynamics is governed by an effective four--dimensional theory for the
zero modes.

Of course, the early universe is not well enough understood to be able to
assert that it must look four--dimensional by the end of inflation. A
conservative estimate would say that
nucleosynthesis~\cite{nucleo1}-\cite{nucleo2} is the earliest time at
which we may be sure standard cosmology applies. However, in part
motivated by the energy scales of Ho\v{r}ava--Witten theory, in this
paper we will indeed look for inflationary solutions that do evolve
from a five-- to a four--dimensional regime during inflation. This has
the advantage of allowing us to follow an explicit cosmological
solution to the full five--dimensional theory as it evolves, and then
match it on to standard cosmology once a 4D description applies. It is
only when one has such an explicit background that it is possible to
apply the formalism developed in
Refs.~\cite{Mukohyama:2000ui}--\cite{vandeBruck:2000ju} to
describe the evolution of cosmological fluctuations during the
five--dimensional era and predict how they would appear in the CMB
today.

Our search for such backgrounds will proceed in two steps. Based on the
classification of solutions, we will first analyze how the properties
of a given class change as we vary the constant potentials $V$ and
$V_i$ ``by hand''. This way, we single out certain classes which are
generically five--dimensional but develop an effective four-dimensional
description in certain regions of the $V$, $V_i$ parameter space. For
those regions, we will determine the corresponding solution to the
four-dimensional effective action.  Of course, since the universe is
dynamic, the mere existence of such regions of parameter space is
insufficient. For the solutions to be realistic, it is also necessary
that they actually evolve towards this region. Since we wish to know
whether the background solutions dynamically evolve towards a
four--dimensional description, it is obviously not appropriate to
study slow--roll within a four--dimensional context from the start. We
will therefore apply our results for five-dimensional slow-roll to
single out appropriate backgrounds. As an additional complication,
it is necessary to ensure that slow--roll does not attempt to take
the full solution out of the region of parameter space for which
it is valid.

\vspace{0.4cm}

Let us now summarize our main results. Our classification of $(A)dS_5$
background solutions leads us to consider four different types of
solutions specified by the signs of $V$ and $V_1+V_2$ which turn out
to have quite different properties. We explicitly determine all
solutions for these four classes. Most notably, we show that for the
case $V>0$ and $V_1+V_2>0$ no non--singular static orbifold solution or
separating solution exists. However, we do find non-separating
solutions with dynamical orbifold in this case. To our knowledge,
these are the first $(A)dS_5$ solutions which allow one to have all
potentials positive, that is $V>0$ and $V_i>0$ where $i=1,2$.  Under
certain additional assumptions, we show that similar statements
hold for the case $V<0$, $V_1+V_2>0$. Another general results concerns
the ratios $\r_i\equiv V_i/\sqrt{12|V|}$. We show that solutions with
$V<0$ only exist if the boundary potentials satisfy the additional
constraints $|\r_i|\geq 1$ and $V_1V_2<0$. No such constraints apply
to the case $V>0$ where $V_1$ and $V_2$ can be chosen arbitrarily.

Which of these solutions exhibit a four--dimensional limit with respect
to the flat vacuum state? As a general rule of thumb, we find that
such a limit can only be realized in the region of parameter space
characterized by $|\r_i|\ll 1$. This basically excludes all solutions
with $V<0$ as they require $|\r_i|\geq 1$. Further, for solutions
with $V>0$, being in the region where $|\r_i|\ll 1$ is a sufficient
condition for exhibiting a $D=4$ limit as long as the orbifold is
static. For a dynamical orbifold, on the other hand, having
$|\r_i|\ll 1$ often leads to a four--dimensional limit only
within a certain time range. This is precisely what happens for
solutions with $V>0$ and $V_1+V_2>0$ which, as we have mentioned,
always have a dynamical orbifold. However, within the class $V>0$
and $V_1+V_2<0$, we do find examples exhibiting a $D=4$ limit whenever
$|\r_i|\ll 1$. For each of these solutions, we examine
the regime where they possess a four--dimensional description, and
explicitly calculate the four--dimensional metric.
For static orbifold cases, the four--dimensional limit is simply $dS_4$
while for dynamical orbifold we typically find power--law expansion.

\vspace{0.4cm}

In order to study how these models evolve during inflation, we have
undertaken a thorough investigation of the slow--roll conditions that
should be imposed on the fields to make the background solutions
consistent. For the bulk scalar field $\f$, this task is
significantly more complicated than in the standard four--dimensional
case. This is partly because the form of the background metrics means
it is impossible to consider $\f$ as independent of the orbifold
co-ordinate and partly because there are boundary conditions
(analogous to the Israel conditions on the metric) that $\f$ must
satisfy at each of the branes. We have found the appropriate
slow--roll equation for $\f$ and its general bulk solution. The form
of this is somewhat complicated and hence we find explicit solutions
to the boundary conditions in two cases; when the bulk and brane
potentials are related by $V_i = U_i(\f_i)[V(\f)]^{\frac{1}{2}}$ and
for arbitrary potentials where the background metric has a static
orbifold radius.

\vspace{0.4cm}

Finally, we have combined the above results to study the dynamical
transition from five to four dimensions. From our statements above
it is clear that this involves having the scalar fields slow--roll
such that the quantities $|\r_i|$ evolve from values
$|\r_i|\geq 1$ at the beginning of inflation to $|\r_i|\ll 1$ at the end.
Seemingly, the solution with all potentials positive is the most
attractive to dynamically realize such an evolution. We find that this
solutions indeed rolls towards the Minkowski vacuum. Unfortunately,
because of the expanding orbifold, the solution becomes again
five--dimensional after a certain critical time corresponding to
a few e-folds, at most. Therefore, for this solution, most of inflation
takes place in five-dimensions and the transition to $D=4$ must happen
during either (p)reheating or radiation domination.
At present we do not possess explicit solutions to the
five--dimensional equations during these epochs, and so for now we
must set this solution to one side.

This leaves us with the solutions for $V>0$ and $V_1+V_2<0$.  At
first, it seems impossible that slow--roll allows us to evolve close
to the Minkowski vacuum at all in those cases -- how can the negative
brane potential which is required roll ``uphill''? We show that this
can happen so long as the brane potential depends on the bulk scalar
field $\f$ more strongly than $\sqrt{V(\f)}$ and less strongly than
$V(\f)^2$. If this is true, then there is no objection to a
four--dimensional description at the end of inflation. This leads to
one of the main results of this paper -- we present a background
solution that undergoes arbitrarily many $e$--folds of
five--dimensional inflation and is slow--rolling to a
four--dimensional description before inflation ends. However, we also
point out that there remains a problem with the stability of the
background solution which is closely related to the problem of how to
stabilize the orbifold. As a concrete application, we demonstrate how
the transition occurs for two simple choices of bulk scalar potential,
a ``new inflation'' model with $V(\f) = M^2 - \frac{1}{2}m^2\f^2 +
\cdots$ and a ``chaotic'' model with $V(\f) = \frac{1}{2}m^2\f^2$. The
purpose here is not to claim that these potentials are realistic, but
simply to show explicitly how the transition to four dimensions
occurs. Interestingly, for the two potentials, the transition occurs
at different rates, and at different stages of inflation. It is
conceivable that still different behaviour would be exhibited by other
potentials, such as those inspired by supersymmetry~\cite{Lyth}.

After this transition, the solutions behave as either $dS_4$ or
power--law inflation, depending on the behaviour of the orbifold
radius. For the first time, this solution allows one to make contact
between the physics of perturbations in brane--world models and
standard four--dimensional cosmology.

The paper is organized as follows. In section~\ref{theory} we review
the five--dimensional action which constitutes the starting point for
our investigation. This may be familiar to many readers and is
included for completeness. Section~\ref{4d} outlines the procedure for
obtaining both the four--dimensional theory based on the Minkowski
vacuum and the general conditions for a five-dimensional solution to
have an effective four-dimensional limit. In section~\ref{sol} we
categorize all the five--dimensional solutions with $(A)dS_5$ bulk
geometry according to the signs of their potentials. The calculations
underlying this categorization are somewhat technical and can be found
in Appendices~\ref{app:cg}-\ref{app:gg}. Next, we focus our attention
on the solutions with $V>0$ and derive their four--dimensional limits
in accordance with the general theory of section~\ref{4d}. In
section~\ref{slowroll} our attention shifts to establishing the
general slow--roll conditions. Section~\ref{applic} then discusses how
each of the $V>0$ solutions evolves under slow--roll and whether it is
possible for them to achieve $\sim 65$ $e$--folds of inflation
combined with a transition to $D=4$. Finally, in section~\ref{phen} we
study the behaviour of a certain class of solutions that has met all
the criteria, showing explicitly how the four--dimensional transition
occurs. This final section is somewhat self--contained and includes a
brief review of the main results of the rest of the paper for those
readers who may be more interested in brane--world phenomenology then
technical details.


\section{The Theories}
\label{theory}

This section will briefly review the five dimensional theory, first
proposed in Ref.~\cite{binf}, that forms the basis for our
investigation of cosmology in the presence of branes. After specifying
the vacuum states of this theory, we derive the effective
four--dimensional theory associated with the flat vacuum state. This
effective action forms the low energy limit of the theory which one
expects to govern the late ``standard'' evolution of the universe.

\subsection{The five--dimensional action}
\label{5d}

The action for our five dimensional theory is given by
\begin{equation}
S_5 = - \frac{1}{2\kappa^{2}_{5}} \left[ \int_{\mathcal{M}_5}
      \sqrt{-g}\left[ R + \frac{1}{2} \partial_{\alpha}\phi
      \partial^{\alpha}\phi + V(\phi)\right] + \sum_{i=1}^{2}
      \int_{\mathcal{M}_{4}^{i}} \sqrt{-g} \left[ \frac{1}{2}
      \partial_{\mu}\phi_{i}\partial^{\mu}\phi_{i} + V_{i}(\phi_{i},
      \phi_i)\right]\right]. \label{action}
\end{equation}
Co-ordinates $x^{\a}$ with indices $\alpha, \beta, \gamma, \ldots =
0,\ldots,3,5$ label the bulk space $\mathcal{M}_5$. The fifth
dimension is compactified on a circle with coordinates $y\equiv x^5\in
[-R,R]$ and the endpoints of the interval identified. As usual, the
orbifold $S^1/Z_2$ is obtained by implementing the $Z_2$ action
$y\rightarrow -y$.  The branes ${\cal M}_4^i$, $i=1,2$ are located at
the fixed points $y=y_i=$ const of the orbifold, where $y_1=0$ and
$y_2=R$. Four-dimensional coordinates longitudinal to the brane will
be denoted by $x^{\m}$ with $\mu, \nu, \ldots = 0, \ldots ,3$. With
our normalization, the brane scalar fields and potentials have mass
dimensions $-\frac{1}{2}$ and 1 respectively, whereas the bulk scalar
fields and potentials have mass dimensions 0 and 2.  The above action
is motivated by five--dimensional heterotic
M-theory~\cite{Lukas:1999yy} and is designed to capture the main
properties of this theory which are essential for cosmology.  Within
heterotic M--theory, the potentials $V$ and $V_i$ have certain
universal parts depending on the dilaton which we have not made
explicit, for simplicity. Also, there are further perturbative
contributions to $V_i$ which are, in principle calculable but rather
model-dependent. Some progress has been made in
Ref.~\cite{Lukas:1999kt,Harvey:1999as,Lima:2001jc} towards
understanding the bulk potential $V$ which, apart from its universal
piece, is of non--perturbative origin. In view of the
model--dependence of all these considerations, here we will view $V$
and $V_i$ as arbitrary, and look to specify what properties they
should have in order to be cosmologically viable. Notice in particular
that the brane scalar field potential may in general depend on both
the brane and bulk scalar fields.

We take the most general Ansatz for the metric consistent with a
homogeneous, isotropic, three--dimensional universe with the
additional simplification of spatial flatness as we expect to be
appropriate during inflation\footnote{For certain solutions it will be
trivial to drop this restriction.}. Our metric Ansatz is then given by
\begin{equation}
 ds^{2}_{(5)} = -e^{2\nu(\tau,y)}d\tau^2 + e^{2\alpha(\tau,y)}
                \d_{ij}dx^{i}dx^{j} + e^{2\beta(\tau,y)}dy^{2}\; .
                \label{ansatz}
\end{equation}
Using this Ansatz while varying the above action leads to the
following Einstein equations
\begin{equation}
 3e^{-2\n}(\dot{\a}^2+\dot{\a}\dot{\b})-3e^{-2\b}(\a ''-\a '\b '+2{\a
 '}^2)= \frac{1}{4}e^{-2\n}\dot{\f}^2+\frac{1}{4}e^{-2\b}{\f
 '}^2+\frac{1}{2}V \label{eom00}
\end{equation}
\begin{multline}
e^{-2\n}(2\ddot{\a}+\ddot{\b}+(3\dot{\a}+2\dot{\b}-2\dot{\n})\dot{\a}
+(\dot{\b}-\dot{\n})\dot{\b}) \\-e^{-2\b}(2\a ''+\n ''+(3\a '-2\b
'+2\n ')\a '+(\n '-\b ')\n ') =-\frac{1}{4}e^{-2\n}\dot{\f}^2
+\frac{1}{4}e^{-2\b}{\f '}^2+\frac{1}{2}V
\label{eomij}
\end{multline}
\bea 3e^{-2\n}(\ddot{\a}-\dot{\n}\dot{\a}+2\dot{\a}^2)-3e^{-2\b}({\a
'}^2+\n '\a ') &=&
-\frac{1}{4}e^{-2\n}\dot{\f}^2-\frac{1}{4}e^{-2\b}{\f
'}^2+\frac{1}{2}V \label{eom55}\\ 3(\dot{\a}'+\dot{\a}\a '-\dot{\a}\n
'-\dot{\b}\a ')&=&-\frac{1}{2}\dot{\f}\f ' \label{eom05}  \eea and
scalar field equations in the boundary picture
\begin{equation}
e^{-2\n}(\ddot{\f}+(3\dot{\a}+\dot{\b}-\dot{\n})\dot{\f})- e^{-2\b}(\f
''+(3\a '-\b '+\n ')\f ') = -\pt_\f V \; .
\label{eomf}
\end{equation}
\begin{equation}
\ddot{\f}_i+(3\dot{\a}-\dot{\n})\dot{\f}_i = -e^{2\n}\pt_{\f_i}V_i
\label{eombf}
\end{equation}
Here a dot (prime) denotes differentiation with respect to $\tau$
($y$).  These bulk equations are subject to the boundary conditions
\bea e^{-\b}\a '\left.\right|_{y=y_i} &=&
\mp\frac{1}{12}\left[\frac{1}{2}e^{-2\n} \dot{\f}_i^2+
V_i\right]_{y=y_i} \label{alphab}\\ e^{-\b}\n '\left.\right|_{y=y_i}
&=& \mp\frac{1}{12}\left[-\frac{5}{2}e^{-2\n}
\dot{\f}_i^2+V_i\right]_{y=y_i} \label{nub}\\ e^{-\b}\f
'\left.\right|_{y=y_i} &=&\pm\frac{1}{2}\left[\pt_\f V_i
\right]_{y=y_i}\label{fb} \eea where the upper (lower) sign applies at
the first (second) brane located at $y=0$ ($y=R$). These conditions
describe permissible discontinuities in the gradient of the metric and
bulk scalar field in the $y$--direction caused by their coupling to
the energy--momentum tensors and fields on the branes.

The five--dimensional action \eqref{action} is well--known to admit
two types of vacuum states which couple to constant potentials $V$,
$V_i$ and respect four--dimensional Poincar\'e invariance.  The first
is simply five--dimensional Minkowski space, which requires
$V=V_1=V_2=0$, and the second is the Randall--Sundrum vacuum
\cite{RS1} with metric
\begin{equation}
ds^2_{(5)} = e^{-2ky}\eta_{\m\n}dx^{\m}dx^{\n} + dy^2 \label{RS}
\end{equation}
where $k=\sqrt{-V/12}$, which is a solutions provided
$V_1=-V_2=\pm\sqrt{-12V}$ with our normalisation of the
potentials. Near each of these vacua we have a four--dimensional
effective theory which describes the zero--mode dynamics of the
respective vacuum state. This four--dimensional description is
typically valid as long as all four--dimensional momenta are smaller
than the mass of the first Kaluza--Klein mode of the vacuum state, so
that the excitation of Kaluza--Klein modes is negligible.  In this
paper, we focus on the flat vacuum state which can also be viewed as
an approximation to the domain--wall vacuum of five-dimensional
heterotic M--theory. However, many of the concepts presented in this
paper should analogously apply to the warped vacua of Ref.~\cite{RS1},
although handling the details will be more complicated. It would be
interesting to work this out explicitly.

{}From \eqref{alphab}-\eqref{nub} it is clear that any solution of the
five--dimensional equations of motion will be inhomogeneous in the
orbifold direction whenever the branes' energy--momentum tensors are
non--negligible. This inhomogeneity corresponds to a coherent
excitation of Kaluza--Klein modes. Therefore, from a five--dimensional
viewpoint, smallness of the brane energy-momentum tensors is a
necessary condition for Kaluza--Klein modes to be negligible and,
hence, for the validity of the $D=4$ effective theory.

\subsection{The four--dimensional low--energy description}
\label{4d}

Let us now review the four--dimensional effective action associated to
the flat vacuum state. As always in the presence of branes, the
Kaluza--Klein modes around the vacuum cannot be set exactly to zero in
a consistent way. However, under the conditions stated above, they
constitute small excitations which can be neglected. It is then
appropriate to perform a reduction to $D=4$ using the vacuum solution
\bea ds^2 &=& \bar{g}_{\m\n}dx^\m dx^\n + e^{2\bar{\b}}dy^2 \\ \f  &=&
\bar{\f} \label{zerom} \eea with the slowly-varying collective modes
$\bar{\b}=\bar{\b}(x^\m )$, $\bar{g}_{\m\n} =\bar{g}_{\m\n}(x^\m )$
and $\bar{\f}=\bar{\f}(x^\m )$ inserted. Defining the four-dimensional
fields
\begin{equation}
 g^{(4)}_{\m\n}=e^{-\bar{\b}}\bar{g}_{\m\n}\; ,\qquad T =
 e^{\bar{\b}}\; ,\qquad S = e^{\bar{\f}}\; , \qquad C_i \equiv
 \sqrt{\frac{M_5^3}{3}}\f_i \label{D4fields}
\end{equation}
one finds for the effective action \bea S_{(4)} &=& -\frac{1}{16\p
G_N}\int_{\mathcal{M}_4}\sqrt{-g_4}\left[ R_4+\frac{3}{2}T^{-2}\pt_\m
T\pt^\m T+\frac{1}{2}S^{-2}\pt_\m S \pt^\m S\right]\nn \\ &&
-\int_{\mathcal{M}_4}\sqrt{-g_4}\left[\frac{1}{2}\sum_{i=1}^2
K_i\pt_\m C_i\pt^\m C_i+V_4\right] \label{S4} \eea with the
four--dimensional potential
\begin{equation}
 V_4 = \frac{1}{2T^2}\sum_{i=1}^2V_{4i}+\frac{1}{T}RM_5^3V\label{V4}\;
 .
\end{equation}
Here the boundary potentials $V_{4i}$, normalized to four mass
dimensions, are defined by $V_{4i}=M_5^3V_i$ and the four--dimensional
Newton constant is $G_N = \frac{1}{16\pi RM_5^3}$. The ``K\"ahler
metrics'' $K_i$ are given by $K_i = \frac{3}{2T}$. The above action is
a lowest--order expression and receives corrections from small but
non-vanishing Kaluza--Klein modes which have to be integrated out.
These corrections can be computed explicitly, see for
example~\cite{Lukas:1998fg,Lukas:1999ew}, but they will not be
essential in context of this paper. As usual, the conformal rotation
of the four--dimensional metric performed in eq.~\eqref{D4fields} is
required if the action~\eqref{S4} is to have a canonically normalized
Einstein term. Equivalently, one could work with a Brans--Dicke type
action with the time variation of the Newton constant then controlled
by the modulus $T$. In this paper, we prefer to consider the standard
Einstein--frame action in four--dimensions.

The action \eqref{S4} contains four--dimensional gravity plus various
scalar fields, and represents our model for ``standard cosmology''.

\vspace{0.4cm}

Now that we have presented the five--dimensional theory and its
four--dimensional effective theory in general, let us see how the
relationship works on the level of individual solutions. We start with
an arbitrary solution of the five--dimensional theory, specified by a
certain metric $g_{\a\b}$, a bulk scalar field $\f$ and boundary
scalars $\f_i$. Of course, most of the five--dimensional solutions
cannot be understood as solutions of the four--dimensional effective
theory, simply because they cannot be cast in the
form~\eqref{zerom}. For example, if the brane stress energy is large,
the five--dimensional solution will have a strong dependence on the
orbifold coordinate which prevents it from being ``near'' the flat
vacuum. This means, in particular, that cosmological backgrounds
describing the evolution of the universe may (and typically will)
start in a regime which is genuinely higher-dimensional, that is,
which has no description in terms of the zero-mode action~\eqref{S4}
or any other four-dimensional action with a finite number of fields.
However, under specific conditions a five--dimensional solution may
have a four--dimensional description. How can this be explicitly
tested?  Let us split our five--dimensional solution according to \bea
g_{\a\b} &=& \bar{g}_{\a\b}+\tilde{g}_{\a\b} \label{gsplit}\\ \f &=&
\bar{\f}+\tilde{\f}. \label{fsplit} \eea into bared fields
corresponding to the orbifold average of the full fields and the
$y$--dependent remainder denoted by a tilde. Obviously, this
decomposition can be applied to any $D=5$ solution and the first,
averaged part is automatically in the near--vacuum form~\eqref{zerom}.
To have a four--dimensional description we should then obviously
require that the $y$-dependent part is somehow small. A practical way
of checking this is to insert the decomposition~\eqref{gsplit},
\eqref{fsplit} into the five--dimensional equations of
motion~\eqref{eom00}-\eqref{fb} and to require the terms involving
tilded fields to be negligible compared to terms involving only bared
fields. The remaining, bared terms then basically correspond to the
equations of motion associated to the four-dimensional effective
theory~\eqref{S4} showing that the averaged fields are an approximate
solution of this effective theory. As an example, typical conditions
to be checked are \bea e^{\tilde{\b}} &\ll& 1 \\ \dot{\tilde{\b}}
&\ll& \dot{\bar{\b}}
\label{require}
\eea and similar ones for the metric components. If these conditions
are satisfies a four--dimensional description of the solution in terms
of the effective theory~\eqref{action} is possible. Clearly, for many
five-dimensional solutions one expects those conditions to hold only
in certain regions of the parameter and coordinate space.  For
example, a solution which is genuinely five--dimensional at early time
may evolve to become effectively four--dimensional later on.  It is
precisely this transition between five and four dimensions that we are
after in the present paper. More specifically, we would like to
analyze the possibility of such a transition in the context of
inflationary solutions to our five--dimensional theory~\eqref{action}.


\section{Five--Dimensional Solutions and their Four--Dimensional Limits}
\label{sol}

In this section we will examine cosmological solutions to our
action~\eqref{action} with all scalar fields and, hence, the
potentials $V$, $V_i$ being constants. This is the first step towards
finding inflationary backgrounds. Varying potentials through scalar
field slow-roll will be implemented as a second step later on. Such
cosmological solutions to the action~\eqref{action} coupling to
constant potentials have been first presented in Ref.~\cite{binf} for
the case of vanishing bulk potential. Subsequently, other solutions
with $(A)dS_5$ and $(A)dS_5$--Schwarzschild bulk geometry have been
found, see, for example
Refs.~\cite{Nihei:1999,KK,Binetruy:1999,Ida,KSW,BCG}. Here, we focus
on the class of solutions with $(A)dS_5$ bulk geometry and we will
systematically list all possibilities. Particular emphasis is put on
examining properties which are of relevance for our subsequent
discussion such as existence of static--orbifold solutions and allowed
ranges for the potentials $V$ and $V_i$. Following our general
prescription outlined in the previous section, we will also check the
existence of four-dimensional limits and calculate the corresponding
four-dimensional solutions where appropriate.

\subsection{General properties}
\label{gen}

In order to solve the equations of motion, it is essential to note the
existence of a first integral~\cite{Bin} which, in conformal gauge $\n
=\b$, is given by
\begin{equation}
 \dot{\a}^2-{\a '}^2=\left(\frac{V}{12}+4Ce^{-4\a}\right) e^{2\b}\; .
 \label{first}
\end{equation}
Here, $C$ is an arbitrary integration constant and solutions with
$(A)dS_5$ bulk geometry can be singled out by setting $C=0$.  The
details of the classification are somewhat technical and are,
therefore, provided in the Appendices~\ref{app:cg}--\ref{app:gg}.
Here, we would like to summarize the main results. The structure of
the solution suggests four sub-classes specified by the signs of $V$
and $V_1+V_2$. These sub-classes have quite different properties,
particularly with respect to the existence of static--orbifold
solutions.  The situation concerning such solutions is summarized in
table~\ref{tab:1}. It is particularly remarkable that no non-singular
static--orbifold solution exists in the first case, where $V>0$ and
$V_1+V_2>0$. Hence in case of all potentials being positive, a
situation which seems to be quite attractive in view of inflation, the
orbifold is necessarily evolving in time. It is important to realize
that these results originate from global properties of the embeddings
of the branes in the background $(A)dS_5$ geometry. In particular,
they would be missed if we were to restrict our attention to just one
of the branes. We also demonstrate in Appendix~\ref{app:gg} that
singular solutions with static orbifold and $V>0$, $V_1+V_2>0$
exist. The singularity then typically implies a horizon separating the
two branes.  However, we will not make any explicit use of such exotic
solutions in the following.  A similar no--go theorem excludes the
existence of static--orbifold solutions in the case $V<0$, $V_1+V_2>0$
as long as the metric is restricted to conformal gauge. However, the
obstruction disappears when arbitrary gauge choices are considered
and, although we have not constructed an explicit example, we expect
static--orbifold solutions to exist under those more general
circumstances.

The situation concerning arbitrary solutions with static or non--static
orbifold is presented in table~\ref{tab:2}. We note that there are no
further constraints on the potential values as long as $V>0$ so that
solutions exist for all values $V>0$ and $V_i$
arbitrary. Particularly, we have found solutions for $V>0$,
$V_1+V_2>0$ which are non-separating and have a dynamical orbifold in
accordance with the above no--go theorem. One notes that the situation
is quite different for $V<0$ where the potentials are further
constrained by $|\r_i|\ll 1$ (where $\r_i\equiv V_i/\sqrt{12|V|}$) and
$V_1V_2<0$.
\begin{table}[t]
\begin{center}
\begin{tabular}{|c|c|c|} 
\hline \it{Bulk Potential} & \it{Brane Potentials} & \it{Constraints?}
\\ \hline  $V>0$ & $V_1+V_2 > 0$ & no non--singular solutions \\ $V>0$
& $V_1+V_2 < 0$ & no further constraints  \\ $V<0$ & $V_1+V_2 > 0$ &
no non-singular solutions known \\ $V<0$ & $V_1+V_2 < 0$ & $|V_i| >
\sqrt{12|V|};\;\; V_1 V_2<0$ \\  \hline
\end{tabular}
\caption{\emph{Solutions with a static orbifold radius.}}
\label{tab:1}
\end{center}
\end{table}
\begin{table}
\begin{center}
\begin{tabular}{|c|c|c|}
\hline \it{Bulk Potential} & \it{Brane Potentials} & \it{Constraints?}
\\ \hline $V>0$ & $V_1+V_2 > 0$ & no further constraints \\ $V>0$ &
$V_1+V_2 < 0$ & no further constraints \\ $V<0$ & $V_1+V_2 > 0$ &
$|V_i| > \sqrt{12|V|};\;\; V_1 V_2<0$ \\ $V<0$ & $V_1+V_2 < 0$ &
$|V_i| > \sqrt{12|V|};\;\; V_1 V_2<0$ \\ \hline
\end{tabular}
\caption{\emph{General Solutions.}}
\label{tab:2}
\end{center}
\end{table}
Which of these solutions are of interest for our subsequent
discussion?  As we have mentioned, we would like to consider solutions
which have a chance of evolving ``near'' the flat vacuum state of our
five-dimensional theory. A necessary condition for this to happen is
that a given solution becomes effectively four--dimensional in a
particular region of the parameter space spanned by $V$, $V_i$. In
this section, this will be checked following the procedure described
in section~\ref{4d}. Whether an existing $D=4$ limit, obtained if
potentials are adjusted to a specific range ``by hand'', can actually
be reached dynamically is a different matter and will be analyzed
later in the context of slow--roll evolution.

It seems unlikely, that solutions with $V<0$ will evolve towards a
flat $D=4$ limit and there is, in fact, a formal argument supporting
this.  As is intuitively clear and will be exemplified below, that one
typically needs to be in the limit $|\r_i|\ll 1$ of the parameter
space to have a $D=4$ description for a particular solution. However,
as we have seen, solutions for $V<0$ exist only provided that
$|\r_i|\geq 1$. We, therefore, remain with the solutions for $V>0$
which do exist in the limit $|\r_i|\ll 1$.  During slow--roll, one
expects that the scalar fields will roll downhill, so the criteria
that we wish to end up near a state where all potentials are zero
seems to indicate that we should consider the case where all
potentials are positive initially. These simply dynamical
considerations seems to restrict us further to the case where $V>0$
and $V_1+V_2>0$. In fact we shall see that there is also another,
somewhat surprising possibility, namely that a negative brane
potential may be pushed `uphill' towards zero. This then allows
consideration of a solution where $V>0$ but $V_1+V_2<0$.  Let us,
therefore, discuss the two cases $V>0$, $V_1+V_2>0$ and $V>0$,
$V_1+V_2<0$ separately in some detail.

\subsection{Solution with $V>0$ and $V_1+V_2>0$}
\label{pos}

The solutions of this type are explicitly given in
Appendix~\ref{app:cg2} and repeated here for convenience
\begin{equation}
ds^2_{5} = \frac{1}{f^2} \left\{ \frac{1}{k^2}\x^2 e^{2\x\t}
           (-d\t^2+dy^2)+d{\bf x}^2 \right\}
\label{posmetric}
\end{equation}
where \bea f &=& \s
+e^{\x\t}\cosh\left[(A_1+A_2)\frac{y}{R}-A_1\right] \label{deff} \\
\x&=&\frac{\left|A_1+A_2\right|}{R} \label{xidef} \\ A_i &=& {\rm
arcsinh}(\r_i) \label{Adef} \eea and $k = \sqrt{\frac{V}{12}}$. If we
want $V_1+V_2>0$, then $f$ must be negative in order for the solution
to fit to the boundary conditions, as shown in
Appendix~\ref{app:cg2}. This excludes the choice $\s =0$, which would
lead to a separating solution with static orbifold. Instead the
constant $\s$ must be strictly negative and we have a non--separating
solution with a dynamical orbifold.  Although the
metric~\eqref{posmetric} looks complicated, it is just $dS_5$ as may
be explicitly verified using the coordinate transformations given in
Appendix~\ref{app:gg}. We note, that the solution as given in
Appendix~\ref{app:cg2} leaves the sign of $\x$ ambiguous. Here, we
have focused on the case $\x >0$ since it corresponds to an expanding
three--dimensional universe. The opposite case, $\x <0$, simply
represent the time--reversal of this solution.

For any choice of (negative) $\s$, the metric~\eqref{posmetric}
represents a valid solution whilst $f$ remains negative, that is, for
time values $\t$ satisfying
\begin{equation}
|\x|\t \in \left(-\infty, \ln\left[ \frac{-\s}{\sqrt{1+ \r_{\rm
           max}^2}}\right]\right)
\label{taurange}
\end{equation}
and it becomes singular towards each of these limits. Calculating the
expansion rates of the brane and orbifold scale factors we find \bea
e^{-\n}\dot{\a}\left.\right|_{y=y_i} &=& k\sqrt{1+\r_i^2}
\label{adot1} \\ e^{-\n}\dot{\b} &=&  -k\s e^{-k\t} \label{bdot1} \eea
and therefore the orbifold is expanding more rapidly than the branes
throughout the range of $\t$ for which this solution exists.

There is an elegant geometric interpretation for solutions of
$(A)dS_5$ type. For the case $V<0$ this was first discussed in
Ref.~\cite{KSW} and one can proceed in an analogous way if $V>0$.  For
our example, the branes are $dS_4$ hypersurfaces embedded in a
background $dS_5$, with the bulk being that portion of $dS_5$ which
lies between the two branes. The model may be visualized as two
hyperbolae of curvature $k\sqrt{1+\r_i^2}$ drawn on the surface of a
hyperboloid of curvature $k$ embedded in a six--dimensional Minkowski
space. The initial singularity at $\t=-\infty$ corresponds to an
intersection of these two hyperbolae, from which they then
diverge. The details of this embedding into six-dimensional Minkowski
space are given in Appendix~\ref{app:gg}.

Let us now explicitly carry out the metric--splitting procedure to
check for possible $D=4$ limits,  as described in
section~\ref{theory}.  We write  \bea \a &=& -\ln\left[-g\right] - \ln
\left[\frac{f}{g}\right] \nonumber \\ &=& \bar{\a} + \tilde{\a}\; ,
\label{asplit} \eea and \bea \b &=& -\ln\left[-\frac{k}{\x}\,g
e^{-\x\t}\right] -  \ln \left[\frac{f}{g}\right] \nonumber \\ &=&
\bar{\b} + \tilde{\b} \label{bsplit} \eea where $g$ is defined
as\footnote{There is a subtlety here. Why have we chosen to represent
the cosh by 1 in $g$? Actually, it is possible to replace the cosh by
any number $c$ with $1 \leq c \leq \sqrt{1+\r_{\rm max}^2}$. Such an
alternative replacement would change the details of the following
calculation; in particular there is some freedom in the value of the
critical time $\t_{\rm c}$. However, it will not affect our main
results.}
\begin{equation}
g = \s + e^{\x\t}\; .
\label{gdef}
\end{equation}
In order for the $y$--independent terms $\bar{\a}$ and $\bar{\b}$ to
be a good approximation to the true solution, the equations of motion
show that we should require $\dot{\tilde{\a}} \ll \dot{\bar{\a}}$,
$\ddot{\tilde{\a}} \ll \ddot{\bar{\a}}$, $\dot{\tilde{\b}} \ll
\dot{\bar{\b}}$, $\ddot{\tilde{\b}} \ll \ddot{\bar{\b}}$ and
$e^{\tilde{\b}} \sim 1$. These conditions show that there is an upper
critical time limit prior to which this description must become valid,
given by
\begin{equation}
\t \ll \t_{\rm c} \equiv \frac{1}{\x}\ln\left[\s -\frac{2\s}
       {\sqrt{1+\r_{\rm max}^2}} \right]
\label{tau4d} 
\end{equation}
where $\r_{\max}$ is the larger of the two ratios of the
potentials. This expression only exists if $\r_{\rm max} < \sqrt{3}$,
so there can only ever be a four--dimensional description if both
ratios $|\r_i|$ are less than $\sqrt{3}$. If we want this description
to be valid at late times, then we will in fact need $\r_i \ll 1$.

The interpretation is simple. When $\t \ll \t_{\rm c}$ the solution
admits a four--dimensional description. This is because the
exponential in the denominator of the true solution \eqref{posmetric}
is extremely small when $\t \sim -\infty$ and hence the metric may be
taken as $y$--independent. As $\t$ progresses from $-\infty$, this
exponential becomes larger making the $y$--dependent cosh term more
crucial, until, as $\t \rightarrow \t_{\rm c}$, a four--dimensional
description is no longer possible. This can also be understood in
terms of the geometrical interpretation. At early times, shortly after
the brane--collision, the bulk samples only a small portion of the
$dS_5$ hyperboloid and, therefore, appears approximately flat in the
orbifold direction. As the branes diverge this portion increases and,
after the critical time, becomes so large that the $D=4$ approximation
breaks down.

It will prove useful later to interpret \eqref{tau4d} in terms of
co-moving time on one of the branes. The induced metric on the branes
takes the form
\begin{equation}
ds^2_{{\rm brane},i} = -dt^2 +
\exp\left(2kt\sqrt{1+\r_i^2}\right)d{\bf x}^2
\label{branemet}
\end{equation}
with $t \in (-\ln (-\s ),\infty)$ between the initial and final
singularities. In terms of co-moving time on this brane, a
four--dimensional description of the complete theory is valid until
$t$ approaches a critical time
\begin{equation}
t \rightarrow t_{\rm c} = \frac{1}{k\sqrt{1+\r_i^2}}\ln \left[
                          \frac{1}{\s -\s\sqrt{1+\r_i^2}}\right]
\label{t4d}
\end{equation}
which again only falls in the range of $t$ covered by our solution if
$\r_i < \sqrt{3}$.

If the above constraints are satisfied, then equation~\eqref{D4fields}
shows that this solution has an effective four--dimensional
description with metric
\begin{equation}
ds^2_{4} = -\frac{\x}{k}\frac{1}{\left(\s e^{-\x\t} + 1\right)^3}
           \left(-\frac{\x^2}{k^2}d\t^2 + e^{-2\x\t}d{\bf x}^2\right)
\label{4met1}
\end{equation}
and $T$--modulus
\begin{equation}
T = -\frac{\x}{k}\frac{1}{(\s e^{-\x\t} +1)}\; .
\label{Tmod}
\end{equation}
The conformal factor makes it difficult to express the metric in terms
of four--dimensional co-moving time $t_{4}$ for all $\t$. It is clear,
though, that is does not represent $dS_4$ but is rather conformal to
$dS_4$ due to our Weyl rotation to the $D=4$ Einstein frame. At any
rate, the above metric and $T$ modulus represent an exact solution of
the $D=4$ effective action (with the other scalar fields being
constant), as can be explicitly checked. Further, it is easy to verify
that as $\t \rightarrow 0$ (a limit we may only consider because the
requirement $\r_i \ll 1$ means that the original metric does not
become singular until $\t \rightarrow 0$) the four--dimensional
metric~\eqref{4met1} displays power--law expansion with scale--factor
$\sim t_{4}^3$ in terms of the four--dimensional co-moving time $t_4$.

In summary, we have seen the first example of a $dS_5$ brane--world
solution which displays a four--dimensional limit under certain
conditions.  As expected we found that the parameters $|\r_i|$ have to
be sufficently small for such a four--dimensional interpretation. In
addition, however, this must arise prior tp a certain critical time
beyond which no 4D limit exists. This additional time constraint is
somewhat troubling given that one would like the late universe to be
effectively four--dimensional. Unfortunately, it originates from the
rapid expansion of the orbifold which cannot be avoided as long as
$V>0$ and $V_1+V_2>0$.

\subsection{Solutions with $V>0$, $V_1+V_2<0$}
\label{bneg}

As mentioned earlier, na\"{\i}vely one expects that in order to
slow--roll down towards the Minkowski vacuum, it is necessary to
consider the previous solution where all the potentials were
positive. However, surprisingly it is also possible to approach the
Minkowski vacuum from a state where one or both brane potentials are
negative. We will discuss exactly how this may occur later, but first
let us present the particular background metrics that we wish to
consider and discuss their four--dimensional limits.

The metric here is closely related both to the previous solution
\eqref{posmetric} and to the solutions for $V<0$ presented in
Ref.~\cite{KK}. Again, focusing on the cases with expanding
three--dimensional universe, we choose the solutions with $\x >0$
given in Appendix~\ref{app:cg2}. Explicitly, they read with
\begin{equation}
ds^2_{5} = \frac{1}{f^2} \left\{ \frac{1}{k^2}\x^2 e^{2\x\t}
           (-d\t^2+dy^2)+d{\bf x}^2 \right\}
\label{met}
\end{equation}
with $f$, $\x$, $A_i$ and $k$ as defined in the previous subsection.
For $V_1+V_2<0$, the function $f$ must now be positive as can be seen
from Appendix~\ref{app:cg2}. Hence, unlike for the case $V_1+V_2>0$,
this does not imply any restrictions on the constant $\s$, and for any
choice of $\s$ the metric~\eqref{met} will be valid throughout the
range of $\t$ that is compatible with the requirement $f>0$.  If $\s
\geq 0$, then it is valid for all $\t \in (-\infty,\infty)$. If $\s
<0$, there is a singularity which leads to a semi-infinite time range
\begin{equation}
\x\t \in \left(-\infty, \ln \left[\frac{\sqrt{1+\r_{\rm min}^2}}{|\s|}
      \right]\right)\; .
\end{equation}
Note in particular, that we are now free to choose $\s =0$ which leads
to a separating solution with static orbifold. A co-ordinate
transformation of the form given in Appendix~\ref{app:gg} shows that
this metric also has bulk geometry $dS_5$.

The brane and orbifold Hubble rates are found to be \bea
\left.e^{-\n}\dot{\a}\right|_{y=y_i} &=& k\sqrt{1+\r_i^2}
\label{adot2} \\ e^{-\n}\dot{\b} &=& -\s ke^{k\t} \label{bdot2}\; .
\eea The three-dimensional space is always expanding, while the
orbifold radius expands, contracts or is stationary depending on
whether $\s$ is negative, positive or zero, respectively. In the
stationary case $\s =0$, the physical orbifold radius is easily shown
to be
\begin{equation}
R_{\rm phys} = \left|\frac{2}{k}\sum_{i=1}^{2} {\rm arctan}\tanh
               \left(\frac{A_i}{2}\right)\right|
\label{Rphys}
\end{equation}
Again, we can give a geometrical description along the lines of
Ref.~\cite{KSW}. Realising $dS_5$ as a hyperboloid in six--dimensional
Minkowski space, the $dS_4$ branes can be obtained as intersections of
certain hyperplanes with this hyperboloid. If those two hyperplanes,
or, equivalently, their normal vectors, are parallel we have a static
orbifold solution which corresponds to setting $\s =0$. Conversely,
non-parallel normal vectors indicate a dynamical orbifold and
correspond to solutions with $\s\neq 0$. The mathematical details of
this interpretation are worked out in app.~\ref{app:gg}.

As an aside, we note that for $\s=0$ it is straightforward to
generalize the metric \eqref{met} to include three-dimensional spatial
curvature. One merely replaces $e^{\x\t}$ by $\cosh ( \x\t)$ or $\sinh
(\x\t )$ for positive and negative spatial curvature, respectively. By
doing so it is clear that as $\t$ progresses, three--dimensional
spatial curvature becomes negligible in this solution and working with
flat three--dimensional sections, as we do, is a good approximation.

Let us again explicitly follow the metric splitting procedure outlined
in section~\ref{theory}. This gives \bea \a &=& -\ln \left[ g \right]
- \ln \left[\frac{f}{g}\right] \nonumber \\ &=& \bar{\a} + \tilde{\a}
\label{asplit2}
\eea and \bea \b &=& -\ln\left[\frac{k}{\x}\, g e^{\x\t} \right] - \ln
\left[\frac{f}{g}\right] \nonumber \\ &=& \bar{\b} + \tilde{\b}\; .
\label{bsplit2}
\eea where now g is defined as~\footnote{Again, we could choose a
different constant to replace the cosh.}
\begin{equation}
g = \s + e^{\x\t}\; .
\label{gdef2}\
\end{equation}
The properties of the four--dimensional limits of these solutions
depend critically on $\s$. We will discuss the three cases separately.

\begin{itemize}

\item $\s=0$~: This is the separating case with static orbifold. The
only requirement to be in a $D=4$ limit is simply that $|\r_i| \ll
1$. Due to separability, the higher Kaluza--Klein modes in the
solution, corresponding to the $y$--dependent pieces, do not evolve
relative to the zero modes. Hence, unlike in our previous case, there
is no time limit by when the four--dimensional theory must become
valid. Since the orbifold radius is static, the four--dimensional
metric is just $dS_4$, that is, \bea ds^2_4 &=&
\frac{\x}{k}\left\{-\frac{\x^2}{k^2}d\t^2 + e^{2\x\t}d{\bf
x}^2\right\} \nonumber \\ &=& T\left\{-dt^2_{4} + e^{2kt_{4}}d{\bf
x}^2\right\}  \eea and the $T$--modulus is fixed as a constant
\begin{equation}
T = \frac{\x}{k}\; .\label{Tmod1}
\end{equation}
This represents an exact solution of the $D=4$ effective
action~\eqref{S4}.  Hence, we have seen our first example for a
solution which is genuinely five-dimensional as long as $\r_i\geq 1$
and has an effective four-dimensional description for $\r_i\ll 1$
without any further constraint on time.

\item $\s>0$~: Again there is no time limit on the validity of a
four--dimensional description, the only requirement being $|\r_i| \ll
1$. The four--dimensional metric is now
\begin{equation}
ds^2_4 = \frac{\x}{k} \frac{1}{(1+\s e^{\x\t})^3}
         \left\{-\frac{\x^2}{k^2}d\t^2 + e^{2\x\t}d{\bf x}^2\right\}\;,
\label{4met2}
\end{equation}
with $T$--modulus
\begin{equation}
T = \frac{\x}{k}\frac{1}{(1+\s e^{\x\t})} \;.
\label{Tmod2}
\end{equation}
As before, this constitutes a solution to the $D=4$ effective
action~\eqref{S4}. Again, it is difficult to write the
metric~\eqref{4met2} in terms of four--dimensional co-moving time
$t_{4}$, but once more it is conformal to $dS_4$ with the $T$--modulus
as a conformal factor. At early times when $\s e^{\x\t} \ll 1$, $T$ is
approximately constant, so the metric is close to $dS_4$. As $\t
\rightarrow \infty$ it is clear that $t_4 \sim -e^{-3|\x|\t/2}$ so
that $t_4 \in (-\infty, 0)$. When approaching the upper time limit,
$t_4\rightarrow 0$, the scale-factor and the $T$ modulus contract
according to the power-laws $a_4\sim (-t_4)^{1/3}$ and $T\sim
(-t_4)^{2/3}$. Altogether, this implies that the 4D limit corresponds
to a negative time--branch solution which ends in a four--dimensional
curvature singularity combined with a brane collision.

\item $\s<0$~: Here, the metric~\eqref{met} possesses a singularity
which arises at different finite times $\t$ across the orbifold. Hence
the Kaluza--Klein tower may be expected to have a significant effect
on the validity of a four--dimensional description near this
singularity. As for the solution with all positive potentials, this
leads to a time--limit on the validity of a four--dimensional
description given by
\begin{equation}
\t \ll \t_{\rm c} \equiv \frac{1}{\x}\ln \left[\frac{1}{|\s|}\right]\;
,
\label{tau4d2}
\end{equation}
in addition to requiring that $|\r_i| \ll 1$. In terms of co-moving
time on the branes, which runs from $-\infty$ to $+\infty$, this
four--dimensional description is valid until
\begin{equation}
t \rightarrow t_{\rm c} = \frac{1}{k\sqrt{1+\r_i^2}}\ln \left
                          [ \frac{1}{|\s|(\sqrt{1+\r_i^2}-1)}\right]\;.
\label{t4d2}
\end{equation}
If the above conditions are satisfied, then the effective
four--dimensional metric is the same as the previous case, but with
$\s <0$:
\begin{equation}
ds^2_4 = \frac{\x}{k}\frac{1}{(1-|\s|e^{\x\t})^3}\left\{ -\frac{\x^2}{k^2}d\t^2 + e^{2\x\t}d{\bf x}^2 \right\}
\label{4met3}
\end{equation}
and the $T$--modulus is
\begin{equation}
T = \frac{\x}{k}\frac{1}{(1-|\s| e^{\x\t})}\; .
\label{Tmod3}
\end{equation}
Again this solves the 4D effective action~\eqref{S4}.  As with the
previous case, the metric behaves like $dS_4$ when $|\s|e^{|\x|\t} \ll
1$, and displays power--law expansion at later times. Here, though,
four-dimensional co-moving time is in the range $t_{4} \in
(-\infty,\infty)$ and there is no curvature singularity at finite
co-moving time. Another difference to the previous case is, of course,
that the orbifold is now expanding and the four--dimensional
description breaks down as $\t$ approaches the critical time $\t_{c}$.
\end{itemize}

\begin{figure}
	\centerline{\psfig{figure=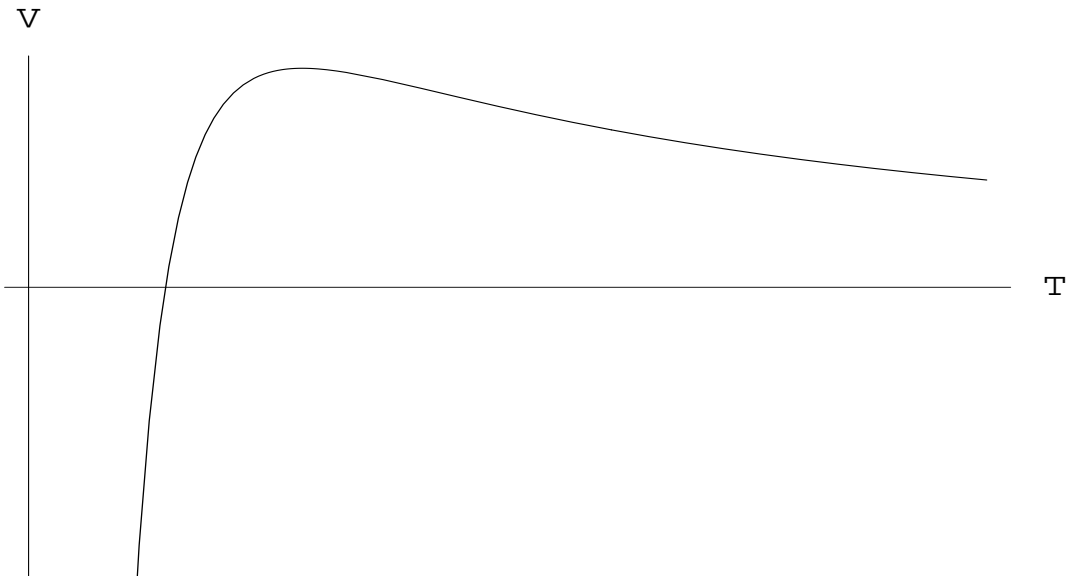,height=3in,width=5.8in}}
	\caption{\emph{Shape of the effective $D=4$ potential for the
	$T$--modulus in the 4D theory resulting from \eqref{met}}.}
	\label{fig1}
\end{figure}

It may help to clarify the situation by discussing the behaviour of
these solutions from the viewpoint of the resulting four--dimensional
action \eqref{S4} and the associated effective potential
\begin{equation}
\frac{V_4}{M_5^3} = \frac{1}{2T^2}\sum_{i=1}^2V_{i}+\frac{1}{T}RV
\label{Tpot}
\end{equation}
for the $T$ modulus. The form of this potential, for $V>0$ and
$V_1+V_2<0$ is schematically shown in figure~\ref{fig1}.  The exact
shape of the potential depends, of course, on the precise values of
$V$ and $V_1+V_2$ but, due to our choices for the signs of these
quantities, it always has a maximum at
\begin{equation}
 T=\frac{V_1+V_2}{RV}\simeq \frac{\x}{k}\; .\label{Tmax}
\end{equation}
The various choices of $\s$ described above can now be understood in
terms of how the $T$--modulus is rolling in this effective potential.
Let us again discuss the three cases separately.
\begin{itemize}

\item $\s=0$~: In this separating case, we have obtained an effective
$D=4$ solution with a constant $T$--modulus. This can only be
compatible with the effective potential~\eqref{Tpot} if $T$ is sitting
at the maximum and a comparison of eq.~\eqref{Tmax} and
eq.~\eqref{Tmod1} shows that this is indeed the case.  This situation
is clearly unstable~\footnote{It is not possible to simply reverse the
signs of all potentials, thus giving the $T$--modulus potential a
minimum. As discussed, in any solution with $V<0$ and $V_1+V_2>0$ one
simply never reaches the four--dimensional regime where a description
in terms of $V_4(T)$ would be appropriate.}. We stress that this
instability, although quite apparent in the four--dimensional limit
was not at all obvious from a purely five-dimensional viewpoint. Of
course, we cannot really tell from this $D=4$ analysis whether the
solution is unstable for any value of $\r_i$ or whether the
instability only arises in the $D=4$ limit when $\r_i\ll 1$. This can
only be clarified by a full five--dimensional stability analysis which
is beyond the scope of the present paper. In any case, our observation
shows that classical stability is an important issue for cosmological
brane--world solutions. To make this solutions realistic, clearly some
additional stabilizing potential, possibly of non--perturbative
origin, has to be invoked. This will be further discussed later on
when slow--roll of scalar fields is analyzed.

\item $\s >0$~: If, either through initial conditions or a
fluctuation, the $T$--modulus were to begin to move towards zero, then
we enter the class of solutions with $\s>0$. As the $T$--modulus
increases its kinetic energy, we enter a power--law expansion as is
usual for exponential potentials ($T=e^{\bar{\b}}$ and $\bar{\b}$ has
the canonically normalized kinetic term). The branes will ultimately
collide at $\t=\infty$ which corresponds to $t_4=0$ in terms of
four--dimensional co-moving time. Realistically, one expects that once
the orbifold radius is smaller than the scale of an underlying
Calabi--Yau space higher--order corrections become important, which
invalidate our simple lowest--order effective actions. One may hope
that these corrections induce a ``graceful exit'' from the negative
time branch, similar to what is normally assumed in pre-Big Bang
cosmology~\cite{pBB1}-\cite{pBB2}.

\item $\s <0$~: The solutions with $\s<0$ correspond to the
$T$--modulus rolling off its maximum in the opposite direction;
$T\rightarrow \infty$. Because $V(T)$ is flatter in this direction,
$T$ rolls more slowly, and so co-moving time is now valid for $t \in
(-\infty, \infty)$. However, the four--dimensional description will
break down as $\t \rightarrow \t_{\rm c}$ where the higher modes in
eq.~\eqref{met} become more crucial.
\end{itemize}


\section{The Slow--Roll Approximation}
\label{slowroll}

In the previous section, we identified classes of solutions to the 5D
Einstein equations which possessed a four--dimensional description
around the Minkowski vacuum for some region of their parameter
space. However, these solutions were always treated adiabatically in
the sense that the brane and bulk scalar potentials were varied ``by
hand''. Our next task is to see whether such background solutions do
in fact evolve dynamically towards the four--dimensional regime. As is
usual for inflation, we do not want the evolution of the scalar fields
to appreciably disturb the adiabatic solutions, and so we must impose
slow--roll conditions on the energy--momentum tensors. In this section
we will examine the general slow--roll conditions that should be
imposed on bulk and brane scalar fields for the validity of arbitrary
conformal gauge $(A)dS_5$ background solutions with $dS_4$
branes. This includes all of the solutions of the previous
section. Slow--roll conditions for a brane scalar field have already
been studied in Ref.~\cite{MWBH}. However, apart from considering a
pure bulk cosmological constant, Ref.~\cite{MWBH} looked at the most
general conditions under which accelerated expansion of the brane
scale--factors could occur. Our purpose here is somewhat different. We
ultimately wish to know how the explicit solutions of the previous
section evolve during slow--roll. Therefore we must be sure our brane
slow--roll conditions are appropriate for $dS_4$ branes, and not {\it
e.g.} brane power--law expansion.

For the moment, we will not restrict ourselves to any particular form
for $V(\f)$ or $V_i(\f_i,\f)$, although sometimes we may consider a
special case where they are related. In general, we will see that it
is impossible to completely separate the bulk and brane
conditions. The resulting slow--roll conditions will be applied to our
specific background solutions in section~\ref{applic}.

\subsection{General requirements of slow--roll}
\label{genreq}

For the bulk energy--momemtum tensor to be dominated by the scalar
field potential, the Einstein equations~\eqref{eom00}-\eqref{eom05}
show that one should require \bea \left| e^{-2\n}\dot{\f}^2 +
e^{-2\b}{\f'}^2\right| &\ll&  2|V| \label{bksr1}\\ \left|
e^{-2\n}\dot{\f}^2 - e^{-2\b}{\f'}^2\right| &\ll&  2|V| \label{bksr2}
\eea and in addition
\begin{equation}
\left|\dot{\f} \f '\right| \ll 6\left| \dot{\a}\a '\right|\; .
\label{bksr3}
\end{equation}
The bulk scalar $\f$ has the (conformal gauge) field
equation~\eqref{eomf}:
\begin{equation}
e^{-2\b}(\ddot{\f}+3\dot{\a}\dot{\f})- e^{-2\b}(\f ''+3\a '\f ') =
-\frac{\pt V}{\pt \f}\; .
\label{eomf2}
\end{equation}
Also, $\f$ is subject to boundary conditions at the branes. Since
these boundary conditions would enter the field equation as terms
involving a $\d$--function, it is not possible to neglect them during
slow--roll. Hence we must ensure
\begin{equation}
e^{-\b}\f '\left.\right|_{y=y_i} =\pm\frac{1}{2}\left[\frac{\pt
V_i}{\pt \f}\right]_{y=y_i}\; .
\label{fb2}
\end{equation}

The branes complicate the bulk slow--roll analysis in two
ways. Firstly, for non--zero brane potentials, we cannot take $\f$ to
be $y$--independent, because there would then be nothing to compensate
for the $y$--dependence of the metric components in
equation~\eqref{eomf2}. Hence, because of the branes, even the
slow--roll equation for $\f$ must contain derivatives with respect to
both $y$ and $\t$. Secondly, suppose we were simply to neglect all
second derivative terms in~\eqref{eomf2} by analogy with the standard
case. It is then possible to find the most general solution to the
resulting first order partial differential equation. The only way this
solution can match the boundary conditions~\eqref{fb2} is if
$V_i(\f,\f_i) = f(\f_i)[V(\f)]^{\frac{2}{3}}$. If the brane and
boundary potentials are related in this way, then the explicit
solution for $\f$ shows that it was not consistent to neglect
$(\ddot{\f}-\f '')$ with respect to $3(\dot{\a}\dot{\f} - \a '\f
')$. We conclude that it is not possible to neglect the second
derivatives of the bulk scalar during slow--roll, and again attribute
this effect directly to the presence of the branes.

\vspace{0.4cm}

The energy--momentum tensors of the brane scalar fields $\f_i$ only
appear in the boundary conditions on the Einstein equations
\eqref{alphab}-\eqref{nub}. These show that for $dS_4$ branes one
should require
\begin{equation}
5e^{-2v}\dot{\f}_i^2 \ll 2V_i\; .
\label{bsr}
\end{equation}
The brane scalar fields $\f_i$ each have field equation \eqref{eombf}:
\begin{equation}
e^{-2\n}\left[\ddot{\f}_i+(3\dot{\a}-\dot{\n})\dot{\f}_i\right] =
-\frac{\pt V_i}{\pt \f_i}
\label{bscalar2}
\end{equation}
where all the components of the metric are evaluated at the brane
positions $y=0,R$. This equation has only one independent variable,
$\t$, which makes finding the appropriate slow--roll conditions very
much analogous to the standard case, and certainly much simpler than
the bulk scalar. Let us therefore begin by examining the brane
conditions.

We split the left hand side of \eqref{bscalar2} in a gauge--invariant
way and neglect $\left| e^{-\n}\pt_\t (e^{-\n}\dot{\f_i})\right|$ to
find the slow--roll equation
\begin{equation}
3e^{-2\n}\dot{\a}\,\dot{\f_i} = -\frac{\pt V_i}{\pt \f_i}\; .
\label{fbsr}
\end{equation}
Exactly as expected, \eqref{fbsr} shows that if $e^{-\n}\dot{\a}>0$,
then during slow--roll the brane scalar fields also roll downhill,
irrespective of the sign of their potentials.

If $T_{05}$ is negligible, which we will need to check using the bulk
slow--roll conditions, then we may use the first
integral~\eqref{first}.  For $C=0$ and general time gauge, it be
written as
\begin{equation}
e^{-2\n}(\dot{\a}^2 - {\a '}^2) = \frac{V}{12} \; .
\label{int1}
\end{equation}
We see that the brane Hubble rates are $e^{-\n}\dot{\a} =
\sqrt{\frac{V}{12}+\frac{V_i^1}{144}}$ for any $(A)dS_5$
background. To apply this to slow--roll, we assume that to zeroth
order the metric components in equation~\eqref{fbsr} may be taken to
have their background values as if the potentials were
constant. Equations~\eqref{bsr} and~\eqref{fbsr} then give the brane
$\e$--condition
\begin{equation}
\e_i \equiv
\frac{10}{3V_i}\frac{1}{\left(V+\frac{V_i^2}{12}\right)}\left(\frac{\pt
V_i}{\pt \f_i}\right)^2 \ll 1\; .
\label{epsi}
\end{equation}
This $\epsilon$--condition is appropriate for all of the solutions of
section~\ref{sol}.

As usual, consistency of neglecting the second derivative terms in
\eqref{bscalar2} generates the $\eta$--condition. Consistency of
neglecting these terms gives
\begin{equation}
\left|\frac{4}{3\left(V+\frac{V_i^2}{12}\right)}\frac{\pt V_i}{\pt
\f_i}\frac{\pt^2 V_i}{\pt \f_i^2} +
\frac{e^{-\n}\dot{\f}}{3\sqrt{\frac{V}{12}+\frac{V_i^2}{144}}}\frac{\pt^2
V_i}{\pt\f\,\pt \f_i} \right| \ll \left| \frac{\pt V_i}{\pt
\f_i}\right| \; ,
\label{etc}
\end{equation}
where we have again taken the zeroth order approximation of treating
the background metric as constant. The first term on the lhs of
eq.~\eqref{etc} clearly forms the brane $\eta$--condition and is valid
for all $dS_4$ branes in any background $(A)dS_5$ metric:
\begin{equation}
\eta_i \equiv
\left|\frac{4}{3\left(V+\frac{V_i^2}{12}\right)}\frac{\pt^2 V_i}{\pt
\f_i^2}\right| \ll 1 \; .
\label{etai}
\end{equation}
The second term in eq.~\eqref{etc} takes account of the fact that the
brane potential also depends on $\f$ and must be related to the bulk
slow--roll conditions. Hence we see that in general, slow--roll on the
brane and in the bulk is inextricably linked.

Notice that the brane slow--roll conditions break down when the
potentials approach either $V=V_i=0$ or $V= -V_i^2/12$. This confirms
that slow--roll of the $\f_i$'s will end when the potentials approach
either the Minkowski or Randall--Sundrum values. In addition, as
noticed in refs.~\cite{MWBH}-\cite{CLL}, since $\e_i$ and $\eta_i$ are
more strongly suppressed in the five--dimensional regime than in the
standard case, it is possible for certain types of potential that are
unable to support standard four--dimensional inflation to nonetheless
satisfy \eqref{epsi}. Slow--roll of the brane scalars could then end
naturally if such a model evolved towards a four--dimensional regime,
where the denominator of \eqref{epsi} becomes small, without having to
require any special feature in the potential. We will not investigate
this possibility further here.

\vspace{0.4cm}

Let us now return to the bulk scalar field. As noticed above, it is
not possible to take $\f$ to be $y$--independent, nor to consistently
neglect the second derivatives in its field equation. To be able to
handle these complications, we introduce a function ${\mathcal T}$ by
$\f=\f({\mathcal T}(\t,y))$ where
\begin{equation}
\frac{d\f}{d{\mathcal T}} \equiv -\frac{3}{V}\frac{\pt V}{\pt \f}
\label{Tdef}
\end{equation}
which fixes ${\mathcal T}$ (up to a constant) for any given
potential. Notice that since $\f$ is dimensionless, ${\mathcal T}$ is
also. This transforms the bulk scalar slow--roll field equation to
\begin{equation}
\ddot{{\mathcal T}} + 3\dot{\a}\dot{{\mathcal T}}-{\mathcal T} '' -3\a
'{\mathcal T}' = e^{2\b}\frac{V}{3}
\label{eomT1}
\end{equation}
where in transforming $\ddot{\f} - \f ''$ we have neglected
\begin{equation}
\left|\frac{d^2\f}{d{\mathcal T}^2}\left( \dot{{\mathcal T}}^2 -
{\mathcal T'}^2\right) \right| \ll \left| \frac{d\f}{d{\mathcal
T}}\left(\ddot{{\mathcal T}} - {\mathcal T} ''\right) \right|
\label{etab1}
\end{equation}
which will later form the $\eta$--condition. The solution of
equation~\eqref{eomT1} tells us how ${\mathcal T}$ behaves, which in
turn would describe $\f$ were we to specify a form for $V(\f)$ and
invert \eqref{Tdef}. It is obviously extremely difficult to solve
equation~\eqref{eomT1} exactly, since the metric components $\a$ and
$\b$ also vary due to the slow--roll of the potentials. As a first
approximation, we may treat the potentials in these terms as
constant. Remarkably, it is then possible to find the particular
integral of equation~\eqref{eomT1}. From equation~\eqref{eom_con} in
Appendix~\ref{app:cg} we see directly that the particular integral is
${\mathcal T} = \a$ for arbitrary conformal gauge backgrounds with
$(A)dS_5$ bulk geometry. Clearly, this includes all of the solutions
discussed in section~\ref{sol}

In order to allow ${\mathcal T}$ to match to arbitrary boundary
conditions, one should also consider the homogeneous equation
\begin{equation}
\ddot{{\mathcal T}} + 3\dot{\a}\dot{{\mathcal T}}-{\mathcal T} '' -3\a
'{\mathcal T}' = 0
\label{eomTh}
\end{equation}
so as to obtain the general solution to the slow--roll field
equation. As shown in Appendix~\ref{app:cg}, in conformal gauge the
metric takes the form~\eqref{ds_con}:
\begin{equation}
 ds^2 = \frac{1}{(f_++f_-)^2}\left\{ -\frac{48}{V}f_+'f_-'dx_+dx_-
        +d{\bf x}^2\right\}
\end{equation}
with $f_+$ and $f_-$ functions of $x_+=\t+y$ and $x_-=\t-y$
respectively. It is possible to solve the homogeneous equation by
defining new co-ordinates (canonical ones for $(A)dS$) as \bea X^+ &=&
\sqrt{\frac{48}{|V|}}f_+ \label{tran1} \\ X^- &=&
\sqrt{\frac{48}{|V|}}f_- \label{tran2} \eea and $X^{\pm} = X_0 \pm
X_1$. In terms of these co-ordinates, the homogeneous
equation~\eqref{eomTh} takes the form
\begin{equation}
\frac{\pt^2 {\mathcal T}}{\pt {X_0}^2} - \frac{3}{X_0}\frac{\pt
{\mathcal T}}{\pt X_0} - \frac{\pt^2 {\mathcal T}}{\pt {X_1}^2} = 0
\label{Th}
\end{equation}
for any conformal gauge $(A)dS_5$ background
solution. Equation~\eqref{Th} is now solved by separation of variables
and the result may be expressed in terms of a series of Bessel and
Neumann functions. The general solution to equation~\eqref{eomT1} is
then
\begin{multline}
{\mathcal T}(\t,y) = \a(\t,y) + \left[\l_0^a + \l_0^b
X_0^4\right]\left[\l_0^c + \l_0^d X_1\right] \\ + \sum_{m \neq 0}
X_0^2\left[\l_m^aJ_2(mX_0)+\l_m^bN_2(mX_0)\right]\left[\l_m^c\sin
(mX_1) + \l_m^d\cos (mX_1)\right]
\label{gensol}
\end{multline}
where $X_{0,1}$ are the functions of $\t$ and $y$ given in
eqs.~\eqref{tran1}--\eqref{tran2} and $m$ is an arbitrary separation
constant. In principle, the integration constants $\l$ in this series
may be used to satisfy arbitrary boundary conditions on ${\mathcal
T}$. In practise, the transformations~\eqref{tran1}--\eqref{tran2}
deform the shape of the branes such that they do not, in general, lie
along lines of constant $X_0$ or $X_1$. For arbitrary background
metrics, this can make it somewhat complicated to fit the general
solution~\eqref{gensol} to the boundary conditions. Two possible
simplifications immediately present themselves; either we can consider
arbitrary background solutions with a specific relation between the
brane and bulk potentials, or we can keep all the potentials arbitrary
but restrict the class of background solution. Let us consider each of
these cases in turn.

\subsection{Special solution for related bulk and brane potentials}
\label{spec}

In this case we simply take the trivial solution to the homogeneous
equation~\eqref{eomTh} and hence have ${\mathcal T} = \a$. Obviously
this will not be possible for arbitrary $V_i(\f,\f_i)$ and indeed the
boundary conditions~\eqref{fb2} and~\eqref{alphab} show that we must
choose the brane and bulk scalar potentials to be related as
\begin{equation}
V_i(\f,\f_i) = f(\f_i)\, [V(\f)]^{\frac{1}{2}} \; .
\label{specsol}
\end{equation}
However, we do not need to make any new assumptions about our
background. What follows is true for any $(A)dS_5$ bulk in conformal
gauge.

With ${\mathcal T} = \a$, the slow--roll condition~\eqref{bksr1} forms
the bulk $\epsilon$--condition
\begin{equation}
\epsilon \equiv \frac{3}{8V^2}(1+2\r_{\rm max}^2)\left(\frac{\pt
V}{\pt \f}\right)^2 \ll 1
\label{bkeps}
\end{equation}
 and the other two conditions~\eqref{bksr2}--\eqref{bksr3} are always
 less restrictive than this. Equation~\eqref{bkeps} is closely
 parallel to the usual $\epsilon$--condition, with corrections
 depending on the strength of the brane sources. If~\eqref{bkeps} is
 satisfied, then~\eqref{bksr3} shows that $T_{05}$ is negligible,
 providing justification for our use of the first
 integral~\eqref{int1} in deriving the brane $\epsilon$--condition.

The bulk $\eta$--condition comes from consistency of neglecting the
piece of the second derivatives of $\f$ in eq.~\eqref{etab1}. In this
special case we may again use the first integral
equation~\eqref{int1}. This allows us to rewrite the
condition~\eqref{etab1} as
\begin{equation}
\left| \frac{3}{V^2}\left(\frac{\pt V}{\pt \f}\right)^2 -
\frac{3}{V}\frac{\pt^2 V}{\pt\f^2}\right| \ll 1 \; .
\label{etab2}
\end{equation}
The first term in this expression is very similar to the
$\epsilon$--condition. However, unlike the standard case in 4D, it is
not always true that this term will be small whenever the
$\epsilon$--condition is satisfied. In fact, this first term is a more
restrictive $\epsilon$--condition whenever $\r_i^2 < 7/2$. Bearing
this in mind, we have the bulk $\eta$--condition
\begin{equation}
\eta \equiv \left|\frac{3}{V}\frac{\pt^2 V}{\pt \f^2}\right| \ll 1 \;.
\label{etab}
\end{equation}
Finally, the second term in eq.~\eqref{etc} becomes
\begin{equation}
\left|\frac{\pt_\f V}{\pt_{\f_i} V_i} \frac{\pt^2 V_i}{\pt \f \pt
\f_i}\right| \ll 1
\end{equation}

Unlike the brane slow--roll conditions, we see that the bulk
conditions only break down near the Minkowski vacuum, and not the
Randall--Sundrum vacuum. This is to be expected, as the
Randall--Sundrum vacuum is still $AdS_5$ from the bulk point of view.

\subsection{General solution for backgrounds with a static radius}
\label{stat}

For backgrounds where the orbifold radius is static, it is possible to
match the general solution to arbitrary boundary conditions, and so
consider cases with arbitrary potentials. Here we present the
conditions that are required in this case with values of $\r_i \sim
{\mathcal O}(1)$ or smaller. We refer the interested reader to
Appendix~\ref{app:sr} for a derivation and more general results for
arbitrary $\r_i$.

The brane $\epsilon$ and $\eta$ conditions are again given by
equations~\eqref{epsi} and~\eqref{etai}. This is to be expected as no
assumptions were made about the background in deriving these
conditions. The bulk field now has a slow--roll equation
\begin{equation}
e^{-\n} \dot{\f} = - 4\sqrt{\frac{V}{12}} \left(\frac{1}{V}\frac{\pt
V}{\pt \f} - \frac{1}{2(V_1+V_2)}\frac{\pt}{\pt \f}(V_1+V_2)\right)\; .
\label{bsrsig0}
\end{equation}
Inserting this into the requirements~\eqref{bksr1}--\eqref{bksr3}
gives the conditions \bea \epsilon &\equiv&
\frac{2}{3V^2}\left(\frac{\pt V}{\pt \f}\right)^2 \ll 1 \label{bkeps1}
\\ \epsilon_{12} &\equiv& \frac{1}{24(V_1+V_2)^2}\left(\frac{\pt}{\pt
\f} (V_1+V_2)\right)^2 \ll 1 \label{bkeps2} \eea The first of these
conditions ensures slow--roll of the bulk scalar in its own potential,
and is closely analogous to the standard condition. The second ensures
that the evolution of the bulk scalar does not cause the brane
potentials to come out of slow--roll. Clearly, each of these should
possess a corresponding $\eta$--condition. The $\eta$--condition for
the bulk field alone is again found by consistency of
eq.~\eqref{etab1} and requires
\begin{equation}
\eta \equiv \left| \frac{3}{V}\frac{\pt^2 V}{\pt \f^2}\right| \ll 1
\label{bketa}
\end{equation}
exactly as in eq.~\eqref{etab} for arbitrary backgrounds with related
brane and bulk potentials. This is again expected. The
$\eta$--condition for the bulk field alone does not notice the
presence of the branes. An $\eta$--condition corresponding to
condition~\eqref{bkeps2} should also be imposed in principle. This is
once more complicated, and here we simply remark that potentials which
satisfy all the other conditions are also likely to satisfy this.


\section{Application to the Background Solutions}
\label{applic}

The formalism of the previous section has introduced slow--roll
parameters $\epsilon$ and $\eta$ for the bulk and brane scalar
fields. In addition, we have confirmed the intuition that, during $dS$
inflation, scalar fields roll downhill towards lower values of their
potentials. We now wish to apply this to the solutions of
section~\ref{sol}, thus upgrading them from static background spaces
to inflationary models. Following the description in
section~\ref{theory}, we wish to look for solutions that evolve
towards the Minkowski vacuum state with $V=V_i=0$. Notice that if we
were to look for a solution that evolved towards the Randall--Sundrum
vacuum with $V<0$ and $V_1 = -V_2 = \pm \sqrt{12|V|}$, then
na\"{\i}vely we would need to start in a region with the negative
brane potential at a value $|V_{-}| < \sqrt{12|V|}$ so as to roll
downhill towards the vacuum state. Table~\ref{tab:2} shows that there
is no such solution.

How can we decide whether a particular brane--inflationary model is a
viable candidate for cosmology?  This depends on when, during the
evolution of the universe, we envisage that the full five--dimensional
solutions entered the 4D regime. We wish to distinguish three
possibilities:

\begin{itemize}
\item The brane sources are negligible either before inflation begins,
or within a small number of $e$--folds. This could arise because of
some unknown constraint from the underlying heterotic M--Theory, or
simply because of the initial conditions in our patch of the universe.
Since the model is then four--dimensional from the beginning of
inflation, it is unlikely that there are any direct cosmological
signatures of extra--dimensions. In order to detect the
higher--dimensional theory, one would have to hope that it had somehow
left an imprint on the structure of the low--energy effective action.

\item The universe becomes four--dimensional during inflation, but
after a reasonable number of $e$--folds of expansion have occurred. So
long as inflation ended soon after (certainly within another $\sim$ 65
$e$--folds), then a significant fraction of our present Hubble volume
will have originally left the horizon during five--dimensional
inflation. This lays open the possibilities of observable signatures
of the fifth dimension and branes in the inflationary power spectrum
at large scales. In addition, since (p)reheating would occur in the
four--dimensional regime~\cite{reh1}-\cite{reh2}, it is conceivable
that this scenario could avoid constraints (see Ref.~\cite{TMM}) on
the production of incoherent Kaluza--Klein particles.

\item The universe is five--dimensional throughout inflation. Hence
the initial fluctuation spectrum was entirely produced during the
5D regime. At the time of writing, experiments only place strict
constraints on the universe during nucleosynthesis (see
refs.~\cite{nucleo1,nucleo2} for a review), so it is conceivable that
the universe could have remained five--dimensional for part of the
radiation--dominated era. Consequently, the evolution of the initial
fluctuations throughout all this time could be dramatically different
from the standard
case~\cite{Mukohyama:2000ui}--\cite{vandeBruck:2000ju} and hence lead
to striking signatures in the CMB~\cite{MaxiBoom}. Such models are
then of considerable interest, though they may be tightly constrained
by further investigations of (p)reheating~\cite{TMM} or
baryogenesis~\cite{Dvali:1999} in brane--worlds.

\end{itemize}

We will now examine the solutions presented in section~\ref{sol} to
discover into which of these categories they fall. We are primarily
interested in finding models in the second category, because these
both have fluctuations that were formed during the five--dimensional
era now visible in the CMB, and are also explicitly calculable
throughout their evolution.

\subsection{Evolution of the positive potential solution}
\label{posapp}

In section~\ref{sol}, we found that this solution would admit a
four--dimensional description so long as the potentials satisfied
$\r_i < \sqrt{3}$ before time
\begin{equation}
\t_{\rm c} = \frac{1}{|\x|} \ln \left[ \frac{2\s}{\sqrt{1+\r_{\rm
max}^2}}-\s \right]
\label{tau4d3}
\end{equation}
where $\s$ was a positive constant. Let us now make a rough estimate
of the amount of inflation that occurs before this time. Considering
simply the $e$--folds of expansion of the brane scale factors, we have
\begin{equation}
N_e \equiv \int_{t_i}^{t_f} e^{-\n}\dot{\a}\,dt \leq
k\sqrt{1+\r_i^2}\, (t_f - t_i)
\end{equation}
where $t$ is co-moving time measured on one of the branes. The upper
limit comes from assuming that the brane Hubble rate is constant
during inflation, rather than decreasing due to slow--roll. Let us
na\"{\i}vely extrapolate \eqref{posmetric} back to the initial
singularity at $t_i = -\ln \s$ (where the two branes were co-incident
and the full eleven--dimensional theory should really have been used)
and take this as the earliest inflation could begin. Requiring that
inflation ends while the solution still has a four dimensional
description would then give
\begin{equation}
N_e \ll \ln \left[\frac{1}{\sqrt{1+\r_i^2}-1}\right]\; ,
\label{Ne1}
\end{equation}
independent of $\s$. For reasonable ({\it i.e.} not fine--tuned)
values of $\r_i$ this shows that the background
metric~\eqref{posmetric} has at most ${\mathcal O}$({\it few})
$e$--folds of brane expansion before the solution looks
five--dimensional. If we wish to obtain sufficient inflation to solve
the horizon problem, we can be sure that the solution with $V>0$ and
$V_i>0$ will end inflation in the five--dimensional regime. This
argument has ignored the effects of slow--roll, which causes $\r_i$ to
evolve and thus affects $N_e$. However, it is easy to convince oneself
that this will not significantly alter our conclusions.

Hence, if they are realistic at all, solutions where all the
potentials are positive must fall into the third category of
cosmological models; the transition to standard cosmology is made
after inflation has ended. Because of the difficulty of obtaining a
complete (analytic) solution when the branes are
radiation--dominated~\cite{CR,EKR}, and possibly even greater
difficulty of attempting any quasi--realistic form of reheating in
these models, we do not pursue this solution further here.

\subsection{Evolution of the solutions with $V_1+V_2<0$}
\label{bnegapp}

Section~\ref{bneg} presented solutions~\eqref{met} which required
$V_1+V_2<0$.  In section~\ref{slowroll} we proved that, as expected,
brane scalar fields slow--roll towards lower values of their
potentials. Therefore, if we wish to evolve towards the Minkowski
vacuum, it is surprising that we can consider the metric~\eqref{met}
because its negative brane potential appears to need to roll
uphill. It is possible to consider the negative potential as a pure
constant. However, since $V_1+V_2<0$, this will not allow us to reach
the four--dimensional regime during inflation unless we have been
there from the start.

The important observation is that the boundary potentials $V_i$ may
(and in Ho\v{r}ava--Witten theory do) also depend on the bulk scalar
field. For certain choices of $V_i(\f,\f_i)$, the negative brane
potential may indeed roll towards zero through slow--roll of $\f$ down
its own (positive) potential $V(\f)$. To see how this works, remember
that the metrics~\eqref{met} require $|\r_i| \ll 1$ for a
four--dimensional description. In order for $|\r_i|$ to decrease
during slow--roll we require

\begin{equation}
\frac{e^{-\n}\dot{\r_i}}{\r_i} =
\frac{e^{-\n}\dot{\f_i}}{V_i}\frac{\pt V_i}{\pt \f_i} +
\frac{e^{-\n}\dot{\f}}{V_i}\frac{\pt V_i}{\pt \f} -
\frac{e^{-\n}\dot{\f}}{2V}\frac{\pt V}{\pt \f} < 0\; .
\label{decrho}
\end{equation}
If we consider the limiting case where there is no scalar field on the
negative brane\footnote{More realistically, this simply means that any
scalar fields that were present did not possess inflationary initial
conditions, and so quickly became negligible.}, just an extra
contribution to the bulk scalar potential, then we find that this
condition is satisfied so long as $e^{-\n}\dot{\f}<0$ and $V_i$
depends on the bulk scalar more strongly than $\sqrt{V(\f)}$. If $V_i$
also depends on some brane scalar field, then there is an extra
positive term in \eqref{decrho} so it is necessary for the
$\f$--dependence to be correspondingly stronger.

In addition we need to be sure that $e^{-\n}\dot{\f} < 0 $ so that
$\f$ is still rolling downhill! For the separating solution with
$\s=0$, the slow--roll equation~\eqref{bsrsig0} shows $\f$ will indeed
roll downhill so long as $V_i$ depends on $\f$ more {\it weakly} than
$V^2(\f)$. For stronger dependence than this, the brane potentials
will actually also drive the bulk scalar in the ``wrong'' direction.

The particular choice $\s=0$ with potentials related by $\sqrt{V} <
V_i < V^2$ therefore forms an example of the second class of
cosmological models; it undergoes an arbitrarily large number of
$e$--folds of five--dimensional inflation, before becoming
four--dimensional towards the end of inflation when it may be
explicitly matched on to standard inflationary models. It then seems
likely that this would lead to observable signatures in the CMB,
caused by the different behaviour of primordial fluctuations in the
five-- and four--dimensional
eras~\cite{Mukohyama:2000ui}--\cite{vandeBruck:2000ju}.

However, as noticed at the end of section~\ref{bneg}, the model with a
static orbifold is somewhat unstable and in general we would expect
fluctuations to cause us to move into the classes with $\s \neq 0$. If
we move towards positive $\s$ then we have a pre--Big Bang situation,
whereas standard inflation still applies if we move towards $\s <
0$. These solutions could then still be an example of the second class
of cosmological models so long as sufficient inflation can be achieved
before the four--dimensional description breaks down as $\t
\rightarrow \t_{\rm c}$. A similar calculation to the one in
section~\ref{posapp} now shows that the maximum number of $e$--folds
of inflation on the brane before this time is
\begin{equation}
N_e \equiv \int_{t_i}^{t_f} e^{-\n}\dot{\a}\,dt \leq \ln
\left[\frac{1}{|\s|(\sqrt{1+\r_i^2}-1)}\right] - kt_i\sqrt{1+\r_i^2}
\;.
\label{Ne2}
\end{equation}
Since brane time $t \in (-\infty,\infty)$ for this solution, there is
no objection to allowing inflation to start at $t_i \rightarrow
-\infty$ and thus obtaining arbitrarily many $e$--folds of expansion
whilst remaining in a four--dimensional regime.

The above conclusions of course depend on the assumption that the
solutions are still inflating while until after they have become
four--dimensional. In other words, we need to be sure that the
slow--roll conditions can still be satisfied during the transition to
the 4D regime. In addition, we must hope that inflation ends soon
after this transition has occurred; another $\sim$ 65 $e$--folds of
standard inflation would clearly dilute any fluctuations which were
sensitive to the five--dimensional regime. In the next section we will
verify that these additional criteria are met in the context of two
simple models.


\section{Explicit Examples and their Phenomenology}
\label{phen}

In this section, we will focus on cases with a transition from five to
four dimensions during inflation. We will study classes of
inflationary models which realize this transition explicitly and
discuss some of their properties. This, for the first time, opens up
the possibility of connecting genuinely higher--dimensional models of
inflation with the subsequent standard evolution of the universe as
described by a four--dimensional low--energy effective theory. To set
the scene, let us first gather some of the relevant results which we
have obtained so far.

As we have seen, background solutions with all potentials positive, or
more precisely with $V>0$ and $V_1+V_2>0$, do not approach the flat
vacuum state during inflation. For such solutions inflation ends in a
five--dimensional regime with the transition to four dimensions
postponed to a later stage. The seemingly most attractive class of
solutions is therefore not suitable for our purpose.

On the other hand, we clearly need the bulk potential to be positive
if we want to evolve into the flat vacuum state, as we do in this
paper.  What remains are the solutions for $V>0$ and $V_1+V_2<0$
described in section~\ref{bneg}. From our classification in
Appendix~\ref{app:cg2}, they are given by
\begin{equation}
 ds^2 = \frac{1}{f^2}\left\{\frac{12}{V}\x^2e^{2\x\t}(-d\t^2+dy^2)
        +d{\bf x}^2\right\} \label{dsex}
\end{equation}
together with the definitions
\begin{equation}
 f = \s+e^{\x\t}\cosh\left[(A_1+A_2)\frac{y}{R}-A_1\right]\;
 . \label{f1}
\end{equation}
and
\begin{equation}
 A_i = {\rm arcsinh} (\r_i)\; ,\qquad \x
 =\frac{\left|A_1+A_2\right|}{R}\; .
\end{equation}
Here $\s$ is an arbitrary integration constant. We have shown that the
above solutions approach the flat vacuum if $|\r_i|\ll 1$ and if, for
$\s <0$, the additional constraint~\eqref{tau4d2} on the time $\t$ is
satisfied. More precisely, in this $D=4$ limit the $D=5$ background
solutions are well approximated by certain solutions to the
four--dimensional effective action~\eqref{S4}. As shown in
section~\ref{bneg}, the metric for these $D=4$ solutions is given by
\begin{equation}
ds^2_4 = \frac{\x}{k}\frac{1}{(1+\s e^{\x\t})^3}\left\{ -\frac{\x^2}{k^2}d\t^2 + e^{2\x\t}d{\bf x}^2 \right\}
\label{ds42}
\end{equation}
and the $T$--modulus is
\begin{equation}
T = \frac{\x}{k}\frac{1}{(1+\s e^{\x\t})}\; .
\label{T2}
\end{equation}
The value of the constant $\s$ is of particular importance. For $\s
=0$ the five-dimensional solution is separating and has a static
orbifold, a property reflected by a constant $T$--modulus in the
associated four-dimensional solution. Unfortunately, as we have seen,
the constant value for $T$ corresponds to the maximum of the
four-dimensional effective potential~\eqref{Tpot}. Therefore, this
solution is unstable at least once it approaches the $D=4$ limit
$|\r_i|\ll 1$ and will be driven, by small perturbations, to a
solution where $T$ rolls down the potential. This is precisely what is
described by the non--separating backgrounds with $\s\neq 0$. As can
be seen from eq.~\eqref{ds42}, \eqref{T2} their four--dimensional
limits correspond approximately to $dS_4$ with a constant $T$--modulus
before a certain time $\t$ and a power--law expansion with a rolling
$T$--modulus after this time. For a realistic model we should,
therefore, explicitly implement a mechanism to stabilize the orbifold
which will then select a solution close to the one with $\s =0$. We
will not attempt to do this explicitly but adopt a practical approach
and work with background solution for $\s =0$.

\vspace{0.4cm}

We have seen in section~\ref{bneg} that a transition from $D=5$ to
$D=4$ can be implemented ``by hand'' by decreasing $\r_i$ to values
smaller than one and have discussed qualitatively how this transition
can be made to work dynamically. We would now like to realize this
transition explicitly, that is, by quantitatively studying the
slow--roll time--evolution of the various scalar fields in the
model. As explained in section~\ref{slowroll} solving the
five--dimensional slow--roll equations for the bulk scalar field $\f$
is not straightforward and leads to a simple answer only when the bulk
and boundary potentials are related in a particular way. For generic
potentials and the separating background, $\s =0$, the solution has
been explicitly given in Appendix~\ref{app:sr} but it turns out to be
rather complicated. For the present purpose, we will work with the
approximation to this solution discussed in section~\ref{stat} which
holds for sufficiently small $\r_i$.

In this limit, the appropriate slow--roll conditions are given by
equations~\eqref{bkeps1}-\eqref{bketa} for $\f$ and
equations~\eqref{epsi}-\eqref{etai} for $\f_i$. The bulk scalar has
slow--roll field equation~\eqref{bsrsig0} while the brane scalar
satisfies~\eqref{fbsr}.  We remark that a more accurate calculation
based on the results of Appendix~\ref{app:sr} leads to complicated
$\r_i$ dependent corrections in the bulk slow--roll conditions. Since
we plan to study examples where $\r_i$ is at most $\sim {\mathcal
O}(1)$, neglecting these corrections and working with the simple
conditions listed above should be a reasonable approximation which
gives qualitatively correct results.

\vspace{0.4cm}

Note that the rolling of the bulk scalar field is generally due to the
bulk potential as well as due to the boundary potentials, which
results in the two terms in the evolution equation~\eqref{bsrsig0} for
$\f$. Depending on the shape of these potentials this may results in a
complicated evolution. To simplify matters, we will focus on cases
where $\dot{\f}$ does not change sign during inflation, so that $\f$
does not start to roll up its own potential.  Then, the number of
$e$--folds based on the $y$--averaged scale factor $\bar{\a}$ is given by
\bea
 N_e &=& \int_{\f_i}^{\f_f}e^{-\n}\dot{\a}\frac{e^{-\n}d\f}{\dot{\f}} \nonumber \\
 &=& \frac{1}{4}\int_{\f_f}^{\f_i} \left[ \frac{1}{V}\frac{\pt V}{\pt \f} - \frac{1}{2(V_1+V_2)}\frac{\pt}{\pt \f}(V_1+V_2)\right]^{-1} e^{-\n}d\f
\label{N}
\eea
Although expressed as an integral over the bulk scalar field,
evaluating this integral generally requires solving the evolution
equations~\eqref{bsrsig0}, \eqref{fbsr}.

\vspace{0.4cm}

Next, as was considered in section~\ref{phen}, we need to discuss the
qualitative evolution of the potentials. Recall that our background
solution required $V_1+V_2$ to be negative. Hence, at least one
boundary potential needs to be negative. It is clear that downhill
evolution of the corresponding boundary scalar will make this
potential even more negative. This is at odds with our goal to evolve
from order--one values for $|\r_i|$ to values smaller than one. A
resolution has been proposed in section~\ref{bnegapp}. The bulk scalar
field, rolling down its own potential, may reverse the trend if the
$\f$ dependence of the boundary potentials is appropriate. Formally
this is governed by equation~\eqref{decrho}. The first term in the
bracket on the rhs describes the change of $\r_i$ due to the change of
$\f_i$ and is always positive. The second term governs the change of
$\r_i$ due to $\f$. For a decreasing $|\r_i|$ we need the rhs of
eq.~\eqref{decrho} to be negative. By inserting the bulk slow--roll
field equation~\eqref{bsrsig0} we find the conditions stated in
section~\ref{bnegapp} that the boundary potentials $V_i$ need to
depend on $\f$ stronger than $\sqrt{V}$ but weaker than $V^2$.

A simple class of models (although there are many more examples)
satisfying these constraints is specified by a ``separating'' boundary
potential of the form
\begin{equation}
  V_i(\f ,\f_i)=V(\f )U_i(\f_i)
\end{equation}
where $U_i(\f_i)$ are arbitrary potentials for the boundary
scalars. For such potentials we find that equation~\eqref{decrho}
becomes
\begin{equation}
 \frac{e^{-\n}\dot{\r_i}}{\r_i} = - \frac{V}{3U_ie^{-v}\dot{\a}}\left(\frac{\pt U_i}{\pt \f_i}\right)^2 -\sqrt{\frac{V}{12}}\frac{1}{V^2}\left(\frac{\pt V}{\pt \f}\right)^2\;.
\end{equation}

For expanding branes, $e^{-\n}\dot{\a} > 0$ and hence the first term
is positive if $U_i <0$. This will drive $\r_i$ in the wrong direction
{\it i.e.} towards more negative values. It is clear that this can be
overcompensated by the second term if we choose the potential $U_i$
sufficiently flat. This is guaranteed in a particularly simple
limiting case, where the boundary potentials are taken to be
independent of $\f_i$. It is for this specific case that we would like
to analyze two explicit examples. Other more complicated examples can
be discussed along similar lines. We consider positive bulk potentials
$V=V(\f )>0$ and boundary potentials
\begin{equation}
 V_i(\f)=\frac{V(\f )}{M_i}
\end{equation}
with $M_i$ being constant mass scales. To have $V_1+V_2<0$ we require
that
\begin{equation}
 \frac{1}{M_1}+\frac{1}{M_2}<0\; .
\label{Micons}
\end{equation}

We are interested in two different choices for the bulk potential
$V(\f )$, namely an ``inverted quadratic potential''~\cite{Lyth} in
which to implement new inflation, and a monomial potential for chaotic
inflation.

\subsection{New inflation model}

\begin{figure}[t]
	\centerline{\psfig{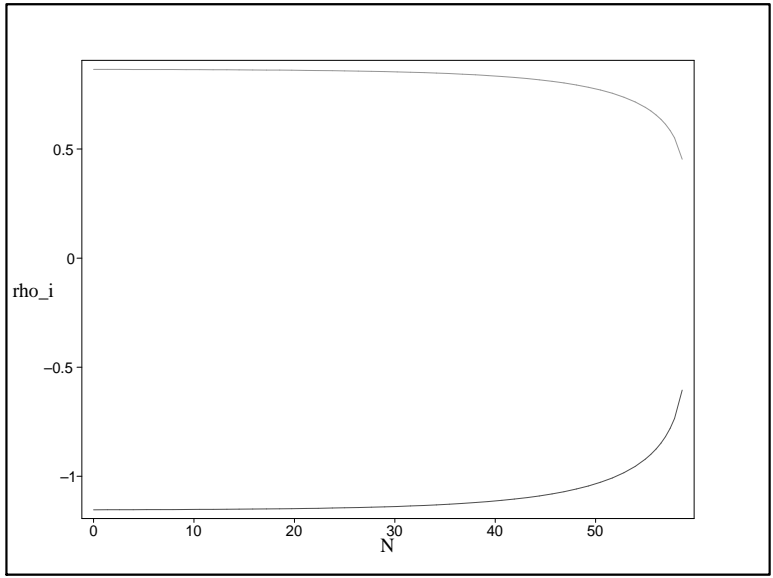}}
	\caption{\emph{Typical evolution of $\r_i=V_i/\sqrt{12V}$ for
	a new inflation bulk potential.}}  \label{fig2}
\end{figure}

We consider the potential
\begin{equation}
 V = M^2-\frac{1}{2}m^2\f^2+\cdots = M^2(1-x^2)+\cdots
\end{equation}
where we have introduced two mass scales $M$ and $m$ with $m\ll M$,
and the rescaled field
\begin{equation}
 x = \frac{m}{\sqrt{2}M}\f\; .
\end{equation}
As usual higher order terms, indicated by dots in the above equation,
have to be added in order to create a minimum. Since we are not
interested in the oscillatory regime here, we need not consider these
terms explicitly. We start inflation at some value $x=x_I\ll 1$ and
roll towards larger values as described by
eq.~\eqref{bsrsig0}. Slow--roll breaks down at $x=x_F\simeq 1$.
Throughout this range $V$ is positive, as is required for our
background solution. The number of $e$--folds $N_e(x)$ between $x_I$ and
$x$ can be obtained from eq.~\eqref{N} and is given by
\begin{equation}
 N_e(x)
 \simeq\frac{M^2}{2m^2}\left(\ln\frac{x}{x_I}-\frac{1}{2}x^2\right)\; .
\end{equation}
This implies for the total number of $e$--folds
\begin{equation}
 N_e \simeq -\frac{M^2}{2m^2}\ln x_I\; .
\end{equation}
As a function of $x$, we find for the all--important quantities $\r_i$
that
\begin{equation}
 \r_i (x) = \frac{M}{\sqrt{12}M_i}\sqrt{1-x^2}
\end{equation} 
which shows that our constraint~\eqref{Micons} on the mass--scales
$M_i$ guarantees a negative value of $\r_1+\r_2$ throughout the
evolution, as is required for the background solution. Most
importantly, we see that both $|\r_i|$ decrease in time. This allows a
dynamical transition from a genuinely 5D regime at $\r_i = {\mathcal
O}(1)$ to an effective $D=4$ regime at $\r_i<1$. To illustrate this
point, we have numerically integrated the evolution equations and, in
fig.~\ref{fig2}, we have plotted $\r_i$ as a function of the number of
$e$--folds, focusing on the last $60$ or so $e$--folds.
As is expected from the form of the potential, $\r_i$ stay
approximately constant for most of the evolution and only change
significantly towards the end of inflation. As a consequence, if one
starts with values of $|\r_i|={\mathcal O}(1)$ at the beginning of
inflation, almost all of inflation takes place in the
five--dimensional regime and the transition to the 4D theory only
happens during the last few $e$--folds.

\subsection{Chaotic inflation model}

\begin{figure}
	\centerline{\psfig{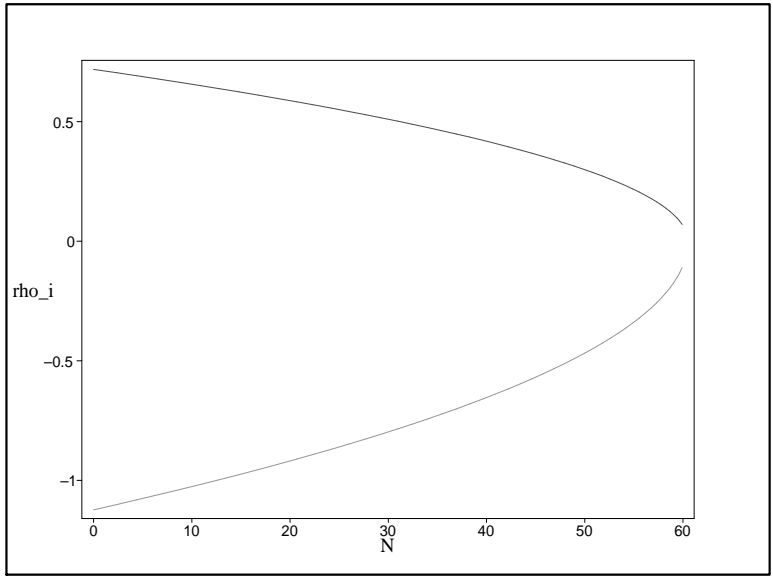}}
	\caption{\emph{Typical evolution of $\r_i=V_i/\sqrt{12V}$ for a 
                       bulk potential of chaotic type.}}
	\label{fig3}
\end{figure}

Let us now contrast this with what happens for a monomial bulk
potential which one expects to be appropriate for chaotic
inflation. We consider
\begin{equation}
 V = \frac{1}{2}m^2\f^2\; .
\end{equation}
We start at some initial $\f = \f_I\gg 1$ and, using eq.~\eqref{bsrsig0}
evolve towards smaller $\f$ with slow--roll breaking down at
$\f = \f_F\simeq 2$. The number of $e$--folds between $\f_I$ and $\f$ is
given by
\begin{equation}
 N_e(\f ) = \frac{1}{8}(\f_I^2-\f^2)
\end{equation}
with the total number of $e$--folds being
\begin{equation}
 N_e = \frac{\f_I^2}{8}\; .
\end{equation}
Furthermore, we find
\begin{equation}
 \r_i (\f ) = \frac{m\f}{2\sqrt{6}M_i}
\end{equation}
so that $\r_1+\r_2<0$ always. Fig.~\ref{fig3} shows the corresponding
plot for $\r_i$ as a function of the number of $e$--folds.
Again, we see that a transition from $D=5$ to $D=4$ can be realized
dynamically. However, unlike the case for new inflation, here the transition starts early on and the system gradually moves towards the four--dimensional regime during inflation.

\vspace{0.4cm}

One expects that the five--dimensional nature of the inflationary backgrounds
just presented will show up in the spectrum of perturbations.
Clearly, the structure of $D=5$ brane--world scalar perturbations and
their evolution equations is much richer than in $D=4$, a fact which
supports this view. Moreover, one hopes that the different types
of transitions to the four--dimensional regime which we have just encountered will lead to different imprints on the spectrum. A reliable analysis of these
questions is beyond the scope of the present paper, and must to be
carried out using the formalism for $D=5$ brane--world perturbations
as, for example, developed in
Refs.~\cite{Mukohyama:2000ui}--\cite{vandeBruck:2000ju}.


\vspace{1cm}

\noindent
{\Large\bf Acknowledgments}

The authors would like to thank Graham Ross, Subir Sarkar and Ed Copeland
for helpful discussions. D.~S.~is supported by a PPARC studentship.
A.~L.~is supported by a PPARC Advanced Fellowship.

\newpage


\vspace{1cm}

\appendix{\noindent\Large\bf Appendix}

\section{Classification of $dS_5$ and $AdS_5$ Solutions in Conformal Gauge}

\label{app:cg}

In this Appendix, we will be systematically analyzing the solutions to
the equations of motion~\eqref{eom00}--\eqref{eom05} which can be
obtained in conformal gauge $\nu = \b$ and which are part of $dS_5$ or
$AdS_5$. Here and throughout the paper, we will always take the branes
to be located at $y=y_i=$ const where $y_1=0$ and $y_2=R$. This
choice, of course, implies that gauge--equivalent solutions to the
bulk equations of motion may not be equivalent with respect to these
fixed boundary conditions.  Therefore, in order to find all possible
solutions, we should  aim to incorporate co-ordinate
reparametrisations into the bulk solution before matching to the
boundaries\footnote{Alternatively, one may incorporate this freedom by
working with arbitrary embeddings for the branes, a route that has
been taken, for example, in Ref.~\cite{Ida,CR}. However, we will not
pursue this method in the present paper.}. For now, we will focus on
conformal gauge, that is, we incorporate conformal transformations
$x_\pm\rightarrow \tilde{x}_\pm (x_\pm )$ of the light-cone
co-ordinates $x_\pm = \t\pm y$ into the metric. Later, in
Appendix~\ref{app:gg}, we will analyze the most general case by
allowing arbitrary co-ordinate transformations.

For the metric in conformal gauge, that is,
\begin{equation}
 ds^2 = e^{2\b}(-d\t^2 +dy^2)+e^{2\a}d{\bf x}^2
\end{equation}
and $\a$, $\b$ being functions of the light-cone co-ordinates, the bulk
equations of motion read
\bea
 \pt_\pm^2\a -2\pt_\pm\a\pt_\pm\b +\pt_\pm\a^2 &=& 0 \nn\\
 \pt_+\pt_-\a +\pt_+\pt_-\b &=& \frac{V}{24}e^{2\b} \label{eom_con}\\
 \pt_+\pt_-\a +3\pt_+\a\pt_-\a &=& \frac{V}{12}e^{2\b}\nn\; .
\eea
These equations can be solved along the lines of Ref.~\cite{KSW} where
the case $V\leq 0$ has been considered. Following Ref.~\cite{Bin}, one
first notes the existence of a first integral
\begin{equation}
 \pt_+\a\pt_-\a = \left(\frac{V}{48}+Ce^{-2\a}\right) e^{2\b}
\label{fint}
\end{equation}
for the system~\eqref{eom_con} with $C$ being an arbitrary integration
constant. In this paper, we are focusing on the subclass of bulk
solutions with maximal symmetry (that is, $dS_5$ or $AdS_5$ solutions)
which are singled out by choosing $C=0$ as
shown in Ref.~\cite{Ida}. The most general solution is then given by
\begin{equation}
 ds^2 = \frac{1}{(f_++f_-)^2}\left\{ -\frac{48}{V}f_+'f_-'dx_+dx_-
        +d{\bf x}^2\right\} \label{ds_con}
\end{equation}
where $f_\pm = f_\pm (x_\pm )$ are arbitrary functions of the light-cone
co-ordinates and the prime denotes the derivative with respect to
the argument. In order to have the correct signature we should
restrict these functions so that ${\rm sign}(f_+'f_-')={\rm sign}(V)$.
We should now subject these bulk solutions to the boundary conditions
\begin{equation}
 \left. e^{-\b}\pt_y\a\right|_{y=y_i} = 
 \left. e^{-\b}\pt_y\b\right|_{y=y_i} = \mp\frac{V_i}{12}\; .
 \label{bound_con}
\end{equation}
This constrains the functions $f_\pm$ to
\begin{equation}
 f_+(\t ) = F(\t )\; , \qquad f_-(\t ) = {\rm sign}(V)\g_1^2F(\t )+k_1
\end{equation}
where $F$ can be expressed in terms of an arbitrary periodic function
$p$ satisfying $p(\t )=p(\t +2R)$ as
\begin{equation}
 F(t) = e^{-\x\t}p(\t ) +\frac{e^{-\x\t}-1}{e^{-2\x R}-1}K\; .\label{F}
\end{equation} 
The coefficients $\g_i$ and $\x$ are defined by
\bea
|\g_i|&=& \mp s\r_i+s_i\sqrt{{\rm sign}(V)+\r_i^2} \label{gi}\\
\x &=& -\frac{1}{R}\ln\frac{|\g_1|}{|\g_2|} \label{xi}
\eea
where the upper (lower) sign in eq.~\eqref{gi} refers to the first
(second) boundary. Further, $K$ and $k_1$ are arbitrary constants
and we recall that $\r_i$ is given by
\begin{equation}
 \r_i = \frac{V_i}{\sqrt{12|V|}}\; .
\end{equation}
Above, we have introduced a number of signs which will turn out to be
quite relevant in the following. In detail, these signs are given by
\begin{equation}
 s = {\rm sign}(f_++f_-){\rm sign}(f_-') \label{s}
\end{equation}
whereas $s_i=\pm 1$ are arbitrary. Normally, one would expect the sign
$s$ to be uniquely defined throughout spacetime. Changing $s$ requires
the functions $f_++f_-$ or $f_-'$ to vanish for certain $\t $ or
$y$.  This implies the existence of a co-ordinate singularity
corresponding to an infinite scale factor or a horizon,
respectively. Even though the curvature scalar remains finite, those
solutions may be undesirable due to their exotic properties, such as
horizon--separated boundaries or an infinitely large orbifold. Be that
as it may, since $f_\pm$ are left-- and right--movers, one at least
expects that there exists a range of $\t$ for which neither $f_++f_-$
nor $f_-'$ change sign as one moves across the orbifold. If necessary,
we can focus on this range of $\t$ and then $s$ is unambiguously
defined.

We also note that the functions $F$, as specified by eq.~\eqref{F}, can
be characterized as solutions of the generalized periodicity condition
\begin{equation}
 F(\t +2R) = \frac{\g_1^2}{\g_2^2}F(\t ) + K\; .
\end{equation}

\vspace{0.4cm}

We now wish to investigate some properties of the above
solutions. Specifically, we are interested in whether we can find
separating solutions ({\it i.e.} solutions where $\a = \a_0(\t
)+\a_5(y)$ and $\b = \b_0(\t )+\b_5(y)$) and solutions with a
time--independent orbifold ({\it i.e.} solutions with
$\dot{\b}=0$). Firstly, we note that for the above class, it can be
shown that separability of the solution implies a time--independent
orbifold. Secondly, by straightforward computation, one can also
demonstrate that a time-independent orbifold requires the condition
\begin{equation}
 k_1(1+{\rm sign}(V)\g_2^2)= k_2(1+{\rm sign}(V)\g_1^2) \label{static_con}
\end{equation}
to hold, where we have defined
\begin{equation}
 k_1 = {\rm sign}(V)\g_2^2K+k_2\; .
\end{equation}
This relation implies the vanishing of the additive constant
in $f_++f_-$. In view of the structure of the solution~\eqref{ds_con},
this is also what one would have intuitively expected as a necessary
condition for a time--independent orbifold. Using condition~\eqref{static_con}
we can now compute the sign $s$ for a solution with a time--independent
orbifold. Inserting $f_+$ and $f_-$ into eq.~\eqref{s} we find
\begin{equation}
 s = -{\rm sign}(\x )\,{\rm sign}({\rm sign}(V)+\g_1^2) = 
     -{\rm sign}(\x )\,{\rm sign}({\rm sign}(V)+\g_2^2)\; .\label{sval}
\end{equation}

\vspace{0.4cm}

We can distinguish four subclasses within the above set of solutions
corresponding to the four different combinations of the signs of
$V$ and $V_1+V_2$. It turns out that these subclasses have quite
different properties, particularly with respect to separability
and existence of static orbifold solutions. We shall therefore discuss them
separately.
\begin{itemize}
 \item $V>0$~: We first note that the signs $s_i$ appearing in eq.~\eqref{gi}
              have to equal $+1$ for $|\g_i|$ to be well-defined. As a
              consequence, we have 
\begin{equation}
 |\g_i| = \mp s\r_i+\sqrt{1+\r_i^2} \label{gi1}
\end{equation}
Let us now turn to the two sub--cases corresponding to the two different signs
of $V_1+V_2$.
  \begin{enumerate}
     \item $V_1+V_2>0$~: In this case, solutions with a time--independent
       orbifold do not exist. This also implies that there are no separating
       solutions. The proof is as follows. Using $\r_1+\r_2>0$
       we conclude from eq.~\eqref{gi1} that $s|\g_2|>s|\g_1|$ which implies
       a positive value of $s\x$ from eq.~\eqref{xi}. However,
       eq.~\eqref{sval} tells us that $s\x$ should be negative for
       solutions with a static orbifold. It is this sign discrepancy that
       precludes the existence of static orbifold solutions in the case
       under discussion.
     \item $V_1+V_2<0$~: Obviously, an argument analogous to the
       previous case does not lead to any contradiction for $\r_1+\r_2$
       negative. Indeed, separating solutions with constant orbifold
       exist in this case and explicit examples will be given in the
       following Appendix.
   \end{enumerate}
 \item $V<0$~: In this case, the expression~\eqref{gi} for the coefficients
       $\g_i$ becomes
\begin{equation}
 |\g_i| = \mp s\r_i+s_i\sqrt{\r_i^2-1}
\end{equation}
       and one immediately concludes that, for the right-hand side to be
       real, one has to require that
\begin{equation}
 |\r_i|\geq 1\qquad{\rm or}\qquad |V_i|\geq 12|V|\; .
\end{equation}
       In addition, the right-hand side must be positive which leads to
\begin{equation}
 {\rm sign}(\r_1)=-{\rm sign}(\r_2) = -s\; . \label{rsigns}
\end{equation}
       In particular, this implies opposite signs for the boundary potentials
       $V_1$ and $V_2$. Note that these are significant constraints on the
       boundary potentials which arise only if $V<0$. A further difference
       from the case $V>0$ is that
       the signs $s_i$ appearing in eq.~\eqref{gi} are not fixed. We note,
       however, that they can be expressed as
\begin{equation}
 s_i = {\rm sign}(\g_i^2-1)\; . \label{si}
\end{equation}
       As before, we now distinguish the two signs of $V_1+V_2$.
 \begin{enumerate}
  \item $V_1+V_2>0$~: Similarly to the corresponding $V>0$ case,
       no separating solutions or solutions with a static orbifold exist
       in this case. For the proof, let us consider a solution with a
       static orbifold. Then, eq.~\eqref{sval} combined with eq.~\eqref{si}
       tells us that the signs $s_i$ have to be the same on both boundaries
       and are given by $s_i = -s\,{\rm sign}(\x )$. From the definitions
       of $\g_i$ and $\x$, eq.~\eqref{gi} and \eqref{xi}, we also conclude
       that ${\rm sign}(|\r_2|-|\r_1|)=s_i\,{\rm sign}(|\g_2|-|\g_1|)=
       s_i\,{\rm sign}(\x )=-s$. However, this relation is only compatible
       with the sign constraints~\eqref{rsigns} on the boundary potentials
       if $\r_1+\r_2<0$.
   \item $V_1+V_2<0$~: In this case, there is no obstruction to having
       separating solutions and solutions with a static orbifold and we
       will present explicit examples in the following Appendix.
  \end{enumerate}
\end{itemize}

\section{Classification of Solutions with Constant Periodic Functions
         in Conformal Gauge}
\label{app:cg2}

We would now like to make the classification of the previous Appendix
more explicit by focusing on an important subclass of solutions. As
we have seen, the general solution obtained in conformal gauge
depends, among other things, on an arbitrary periodic function
$p$. Physically, this function represents the freedom to impose
arbitrary fluctuations in the orbifold direction as an initial
condition. We remark that such initial fluctuations may not be wiped
out and, hence, constitute a challenge for inflation particularly with
regard to inflationary ``no--hair theorems'' (see~\cite{SFV} for a
recent discussion). Clearly this is a problem present in all
higher--dimensional models of inflation but it is particularly acute in
brane--world models. While we will not attempt to resolve this problem
in the present paper, it is clear that in order to obtain workable
cosmological backgrounds, one would like to concentrate on solutions
where such fluctuations are small or, equivalently, where the
periodic function $p$ is constant. It is this particular subclass that
we now wish to analyze in more detail.

\vspace{0.4cm}

Assuming a constant periodic function $p$ in the eq.~\eqref{ds_con}, it
can be shown that the five--dimensional metric may be written in the
form
\begin{equation}
 ds^2 = \frac{1}{f^2}\left\{ e^{2\b_0-2\x\t}(-d\t^2+dy^2)+e^{2\a_0}d{\bf x}^2
        \right\}\; . \label{ds_con0}
\end{equation}
The function $f$ is defined by
\begin{equation}
 f = \s+e^{-\x\t}\left\{\begin{array}{lll}\cosh (\x y+\x_0)&{\rm for}&V>0\\
                                         \sinh (\x y+\x_0)&{\rm for}&V<0
                       \end{array}\right.
\label{fdef}
\end{equation}
and $\a_0$, $\b_0$, $\x$, $\x_0$ and $\s$ are constants. These constants
can be determined, of course, by comparison with the general
conformal solution~\eqref{ds_con}--\eqref{xi}. However, it is more
instructive, and moreover serves as a cross-check of our previous
results, to insert the metric~\eqref{ds_con0} directly into the
equations of motion and the boundary conditions. One then finds that
the bulk equations of motion~\eqref{eom_con} are satisfied as long
as
\begin{equation}
 |\x | e^{-\b_0} = \sqrt{\frac{|V|}{12}}
\end{equation}
where the sign of $\x$ remains undetermined. To incorporate the boundary
conditions~\eqref{bound_con} it again proves useful to distinguish
four cases as specified by the signs of $V$ and $V_1+V_2$.
\begin{itemize}
 \item $V>0$~: In this case we define
\begin{equation}
 A_i = {\rm arcsinh} (\r_i)
\end{equation}
and the boundary conditions are satisfied if
\begin{equation}
 \x = -{\rm sign}(f\x )\frac{A_1+A_2}{R}\; ,\qquad \x_0 = {\rm sign}(f\x )A_1
 \; . \label{xi1}
\end{equation}
The metric~\eqref{ds_con0} takes the form
\begin{equation}
 ds^2 = \frac{1}{f^2}\left\{\frac{12}{V}\x^2e^{-2\x\t}(-d\t^2+dy^2)
        +e^{2\a_0}d{\bf x}^2\right\}
 \label{ds2}
\end{equation}
where the function $f$ is now given by
\begin{equation}
 f = \s+e^{-\x\t}\cosh\left[(A_1+A_2)\frac{y}{R}-A_1\right]\; . \label{ff1}
\end{equation}
The constants $\a_0$ and $\s$ remain undetermined. The former can
be absorbed into a co-ordinate rescaling and is trivial in this sense.
The constant $\s$, however, cannot be removed in this way and its value
is essential for the properties of the solution. Specifically, the
solution is separating and has a static orbifold if and only if $\s=0$.
In order to see whether such a choice is permissible let us examine
the two signs for $V_1+V_2$ separately.
  \begin{enumerate}
    \item $V_1+V_2>0$~: We easily deduce from the first
     equation~\eqref{xi1} that the function $f$ has to be
     negative. However, from its definition~\eqref{fdef} it is clear
     that this is incompatible with setting $\s=0$. In fact, we need
     $\s<0$. Consequently, no separating solutions or solutions with
     static orbifold exist in accordance with our conclusion in the
     previous Appendix.  As stands, $\t$ takes values in a
     semi-infinite range. For $\x>0$ the solution evolves from a
     co-ordinate singularity at some finite time corresponding to
     infinitely separated branes to brane collision as
     $\t\rightarrow\infty$. The branes scale factors themselves are
     decreasing, so $\x>0$ does not represent inflation. In the
     opposite case, $\x<0$, the models starts out with colliding
     branes at $\t\rightarrow -\infty$ and ends in a co-ordinate
     singularity with infinite brane separation at some finite
     time. Here the branes are inflating. This case is discussed in
     detail in section~\ref{pos}, where $f$ is defined with an extra
     minus sign so as to make it positive.
    
    \item $V_1+V_2<0$~: Eq.~\eqref{xi1} now
     implies that the function $f$ has to be positive. This, however,
     places no further constraint on the constant $\s$. In particular,
     we are free to set $\s=0$ which leads to a separating solution
     with static orbifold~\cite{KK,KSW}.  For $\s \geq 0$ the range of
     $\t$ is infinite, that is,  $\t\rightarrow\pm\infty$ while for
     $\s <0$ it is restricted to  a semi-infinite range with a
     singularity corresponding to an infinite brane-separation
     developing at some finite time. These cases are discussed in
     detail in section~\ref{bneg}.
   \end{enumerate}
  \item $V<0$~: Defining
\begin{equation}
 A_i = {\rm arccosh} (|\r_i|)
\end{equation}
the boundary conditions can be satisfied provided
\begin{equation}
 \x = -\frac{s_1A_1+s_2A_2}{R}\; ,\qquad \x_0 = s_1A_1\label{xi2}
\end{equation}
where $s_i=\pm 1$ are, as yet, undetermined signs. For this to
provide a well-defined solution to the boundary conditions we need
to require in addition that
\begin{equation}
 V_1V_2<0\; ,\qquad |\r_i|\geq 1\; .
\end{equation}
These conditions on the boundary potentials are the same we have found
for the general conformal solution in the previous Appendix, as they
should. The metric is then given by
\begin{equation}
 ds^2 = \frac{1}{f^2}\left\{\frac{12}{-V}\x^2e^{-2\x\t}(-d\t^2+dy^2)
        +e^{2\a_0}d{\bf x}^2\right\}
\end{equation}
with
\begin{equation}
 f = \s+e^{-\x\t}\sinh\left[-(s_1A_1+s_2A_2)\frac{y}{R}+s_1A_1\right]\; .
 \label{f2}
\end{equation}
In fact, our information about the signs of $V_i$ is somewhat more precise,
namely
\begin{equation}
 {\rm sign}(V_1)=-{\rm sign}(V_2)={\rm sign}(f\x )\; . \label{Visign}
\end{equation}
As with the case where $V>0$, the constant $\s$ is significant.
In particular, $\s=0$ characterizes the separating solutions with
static orbifold. To find out whether or not this choice is possible
we again distinguish between the two signs of $V_1+V_2$.
  \begin{enumerate}
   \item $V_1+V_2>0$~: No separating solutions or solutions with
   static orbifold exist in this case. For the proof, let us assume
   the existence of such a solution with $\s=0$. We focus on times $\t$
   where the sign of $f$ in eq.~\eqref{f2} does not change across the
   bulk. Then the signs $s_i$ are fixed to $s_1=-s_2={\rm
   sign}(f)$. From eq.~\eqref{xi2}, one concludes that ${\rm sign}(\x
   ) = {\rm sign}(f(|V_2|-|V_1|))$ and, by comparison with
   eq.~\eqref{Visign}, we learn that ${\rm sign}(V_1)=-{\rm sign}(V_2)=
   {\rm sign}((|V_2|-|V_1|))$. This last equation, however,
   contradicts our assumption that $V_1+V_2>0$.

   The non--separating solutions with $\s\neq 0$ exist with infinite
   or semi--infinite range of $\t$ depending on the sign of $\s$ and
   $s_i$. In either case, the endpoints of the evolution correspond
   to colliding branes or infinitely separated branes.
   \item $V_1+V_2<0$~: In this case, there is no obstruction to setting
   $\s=0$ and, hence, separating solutions with a time--independent orbifold
   exist. The time--range can be infinite or semi--infinite depending on
   the values of $\s$ and $s_i$. For $\s\neq 0$ the endpoints of the
   evolution again correspond to a collapsing or diverging orbifold
   size. These solutions have been discussed in Ref.~\cite{KK}.
  \end{enumerate}
\end{itemize}

\section{Classification of Solutions in Arbitrary Gauge}

\label{app:gg}

As we have mentioned before, the approach of this paper is
to explicitly incorporate co-ordinate reparametrizations into the bulk
metric and keep the boundaries at $y=$ const hypersurfaces rather
than allowing curved branes. So far, we have focused on conformal
transformations and subclasses thereof but it is not clear, {\it a priori},
that this leads to the most general set of solutions. In this Appendix 
we will therefore allow arbitrary co-ordinate transformations.

The corresponding bulk metric is easily obtained from the one in conformal
gauge, eq.~\eqref{ds_con}, by transforming to a new set of co-ordinates,
$(y^a)=(\t, y)$ where $a=0,5$ and $x_\pm = x_\pm (\t ,y)$. We then have
the metric
\begin{equation}
 ds^2 = \frac{1}{(F_++F_-)^2}\left\{\g_{ab}dy^ady^b+d{\bf x}^2\right\}
 \label{ds_gen}
\end{equation}
where
\begin{equation}
 \g_{ab} = -{48}{V}\pt_{(a}F_+\pt_{b)}F_-
\end{equation}
and $F_\pm$ are arbitrary functions of $\t$ and $y$ related to their
conformal gauge counterparts by
\begin{equation}
 F_\pm (\t ,y) = f_\pm (x_\pm (\t ,y))\; .
\end{equation}
In order for this metric to have to correct signature (with $y$ being
the spacelike direction) one should require that
\begin{equation}
 {\rm sign}(V)={\rm sign}(\dot{F_+}\dot{F_-})=-{\rm sign}(F_+'F_-')\; ,
 \label{Fsign}
\end{equation}
at least close to the branes.

In this arbitrary gauge, the boundary conditions for the $(00)$ and
$(ij)$ components of the metric are still given by
eqs.~\eqref{alphab}--\eqref{nub}, while we have to require in addition
that $g_{05}$ vanishes at the branes. Interestingly, these
conditions can be solved explicitly leading to the following
constraints

\begin{equation}
 F_-(\t ,y_i)={\rm sign}(V)\g_i^2F_+(\t ,y_i)+k_i\; ,\qquad
 F_-'(\t ,y_i)=-{\rm sign}(V)\g_i^2F_+'(\t ,y_i) \label{Fbound}
\end{equation}
on the functions $F_\pm$ at the locations $y=y_i$ of the branes.
As before, the co-efficients $\g_i$ are defined by
\begin{equation}
 |\g_i|=\mp s\r_i+s_i\sqrt{\r_i^2+{\rm sign}(V)}\; , \label{gi2}
\end{equation}
where $s_i=\pm 1$ are as yet unspecified signs, $k_i$ are constants
and the sign $s$ is defined by
\begin{equation}
 s = {\rm sign}(F_++F_-)\,{\rm sign}(F_-')\; . \label{sdef}
\end{equation}
Hence, we have found a solution for all functions $F_\pm$ satisfying
the boundary conditions~\eqref{Fbound} together with the additional
sign constraints~\eqref{Fsign}. While these constraints seem
relatively weak at first, they may in fact obstruct the existence of
a solution under certain conditions. Typically, in these cases one
finds that a pair of functions $F_\pm$ which satisfy the boundary
conditions on one side, when continued across the bulk in accordance
with the constraints~\eqref{Fsign} on $F_\pm'$, can never satisfy the
boundary conditions on the opposite side. We will see this explicitly
in a moment.

\vspace{0.4cm}

First, however, it is helpful to discuss some geometrical aspects of our
solutions following the ideas of Ref.~\cite{KSW}. As we will see,
the general approach of this Appendix is rather well--suited for this
purpose. Since our bulk solutions represents a slice of $dS_5$ or
$AdS_5$ they can be represented as a hypersurface
\begin{equation}
 -T^2+Y^2+{\bf X}^2+{\rm sign}(V)Z^2={\rm sign}(V){\cal R}^2\label{hyper}
\end{equation}
in a six--dimensional space with co-ordinates $(T,Y,{\bf X},Z)$ and metric
\begin{equation}
 ds_6^2 = -dT^2+dY^2+d{\bf X}^2+{\rm sign}(V)dZ^2\; .
\end{equation}
Here ${\cal R}$ is the radius of the (anti) de Sitter space, given by
\begin{equation}
 {\cal R}^2=\frac{1}{k^2}=\frac{12}{|V|}\; .
\end{equation}
It is straightforward to show that the six--dimensional embedding co-ordinates
are related to our original five--dimensional ones by
\bea
 T &=& {\cal R}\frac{1-4\,{\rm sign}(V)F_+F_-+{\bf x}^2/{\cal R}^2}
       {2(F_++F_-)} \nn \\
 Y &=& {\cal R}\frac{1+4\,{\rm sign}(V)F_+F_--{\bf x}^2/{\cal R}^2}
       {2(F_++F_-)} \nn \\
 {\bf X} &=& \frac{{\bf x}}{F_++F_-} \label{coordrel}\\
 Z &=& {\cal R}\frac{F_+-F_-}{F_++F_-}\nn \; .
\eea
In the six--dimensional embedding space, the branes are given as the
intersections of certain five--dimensional hyperplanes with the hyperboloid
defined by eq.~\eqref{hyper}. Using the relations~\eqref{coordrel}
together with the boundary conditions~\eqref{Fbound} on $F_\pm$ we can
determine normal vectors ${\bf N}_i$ to these hyperplanes. Up to a
normalisation constant, we find for these vectors
\begin{equation}
 {\bf N}_i \sim (2k_i,2k_i,0,0,0,1+{\rm sign}(V)\g_i^2)\; .
 \label{Ni}
\end{equation}
These vectors carry information about how the branes are embedded into the
$(A)dS_5$ space and, as expected, they depend on the bulk potential V
and the respective brane potential $V_i$ through the co-efficients $\g_i$.
However, we also observe that the constants $k_i$ which appear as integration
constants in the boundary conditions~\eqref{Fbound}, play an important
role in specifying this embedding. We can use the normal vectors ${\bf N}_i$
to distinguish two qualitatively different types of solution. First, if
${\bf N}_1$ and ${\bf N}_2$ are not parallel we have a diverging
orbifold size in the past (future) and colliding branes in the future (past).
In other words, such solutions are characterized by a non--static orbifold.
On the other hand, if ${\bf N}_1$ and ${\bf N}_2$ are parallel the
branes never intersect which constitutes a necessary condition for
the orbifold to be static. From eq.~\eqref{Ni} we see that this case is
realized precisely when
\begin{equation}
 k_1(1+{\rm sign}(V)\g_2^2)=k_2(1+{\rm sign}(V)\g_1^2)
 \label{static}
\end{equation}
which serves as a useful practical condition for a static orbifold.

\vspace{0.4cm}

In order to apply this criterion to our general solution, as before we
distinguish the various cases corresponding to the signs of $V$ and
$V_1+V_2$.
\begin{itemize}
 \item $V>0$~: As in the conformal gauge case, the signs $s_i$ in
  eq.~\eqref{gi1} must equal $+1$ which leads to
\begin{equation}
 |\g_i| = \mp s\r_i+\sqrt{1+\r_i^2} \label{gi3}
\end{equation}
 For the two signs of $V_1+V_2$ we find the following:
  \begin{enumerate}
   \item $V_1+V_2>0$~: The previous no--go theorem excluding the existence of
    solutions with static orbifold can be avoided if the function $F_+'$
    (or $F_-'$) is allowed to vanish at points
    in the orbifold. For such solutions, the metric~\eqref{ds_gen}
    develops a co-ordinate singularity leading to a horizon that separates
    the two boundaries. An example can be obtained by starting
    with the metric~\eqref{ds2} with the constant $\s$ set to zero in
    \eqref{ff1}. If we introduce a new orbifold co-ordinate $z$ by
    $g(z)=1/\cosh(\x y+\x_0)$, where $g$ is an arbitrary function
    the transformed metric reads
\begin{equation}
 ds^2 = -\frac{12\x^2}{V}g(z)^2d\t^2+\frac{12}{V}\frac{g'(z)}{1+g(z)^2}dz^2
        +e^{2\a_0+\x\t}g(z)^2d{\bf x}^2\; .\label{ds_counter}
\end{equation}
    We find that the boundary conditions~\eqref{alphab}--\eqref{nub}
    are solved in these new co-ordinates if
\begin{equation}
 {\rm sign}(g(z_i))\sqrt{1+1/g(z_i)^2}=\mp\r_i \label{gi4}
\end{equation}
    If $g(z)\neq 0$ throughout the orbifold, the signs of $g(z_i)$ must
    be the same and the above conditions cannot be satisfied on both
    boundaries given that $\r_1+\r_2>0$. On the other hand, if $g(z_1)$
    and $g(z_2)$ have different signs the obstruction disappears and
    functions $g$ matching to both boundaries can be found. The function
    $g$ must then vanish at at least one point in the orbifold. This results
    in a co-ordinate singularity in the metric~\eqref{ds_counter} associated
    with a horizon separating the two boundaries. An analogous example,
    for the case of a flat bulk spacetime, has been found in
    Ref.~\cite{binf}.

    Having presented the above counter--example, let us now exclude
    co-ordinate singularities and assume that the functions $F_+'$,
    $F_-'$ as well as $F_++F_-$ are non--vanishing throughout the
    spacetime. Under this additional assumption, it can again be
    shown that static orbifold solutions do not exist. For the proof,
    we have to analyze the boundary conditions on $F_\pm$ given by
    eq.~\eqref{Fbound}. A useful observation which simplifies the
    argument is that these functions can be shifted by a constant $c$
    according to $F_+\rightarrow F_+-c$ and $F_-\rightarrow F_-+c$
    without affecting the bulk metric. The boundary conditions will,
    of course, be modified. Let us choose the constant $c$ to be
    $c=k_1/(1+\g_1^2)$. After the shift the additive constant in the
    condition on the first boundary will be removed. If, in addition,
    we assume a static orbifold, we learn from eq~\eqref{static} that
    the same is true for the second boundary. Then, the new boundary
    conditions are simply
\begin{equation}
 F_-(\t ,y_i)=\g_i^2F_+(\t ,y_i)\; .\label{Fbound1}
\end{equation}
    These equations, together with our initial assumption about the
    absence of co-ordinate singularities implies that the signs
    of $F_+(\t ,y_i)$ and $F_-(\t ,y_j)$ must be the same for all $i,j=1,2$.
    In addition, when continued across the orbifold the functions
    $F_\pm$ have to respect the sign constraint~\eqref{sdef}. This
    leads to $s|F_-(\t ,y_1)|>s|F_-(\t ,y_2)|$ and 
    $s|F_+(\t ,y_1)|<s|F_+(\t ,y_2)|$. From eq.~\eqref{Fbound1} this
    means that we can only find functions $F_\pm$ matching both
    boundaries if $s\g_1^2>s\g_2^2$. However, this contradicts the
    definition for $\g_i$, eq.~\eqref{gi4} which implies that
    $s\g_1^2<s\g_2^2$ as long as $\r_1+\r_2>0$.
   \item $V_1+V_2<0$~: As we have already seen, static orbifold solutions
   exist in this case.
  \end{enumerate}
  \item $V<0$~: The coefficients $\g_i$ are given by
\begin{equation}
 |\g_i| = \mp s\r_i+s_i\sqrt{\r_i^2-1} \label{gi5}
\end{equation}
   and, as in the more special cases, are only well--defined if
   the potentials obey
\begin{equation}
 |V_i|\geq 12|V|\; ,\qquad V_1V_2<0\; .
\end{equation}
   Hence, these significant constraints on the potentials persist even
   in the most general case. If they are violated no solutions exist.
 \begin{enumerate}
  \item $V_1+V_2>0$~: Arguments analogous to the ones in the $V>0$ case do not
  apply any longer, essentially due to the additional sign freedom
  $s_i$ in eq.~\eqref{gi4} which, unlike in the case of conformal gauge,
  can no longer be fixed. We expect static orbifold solutions to
  exist, even if the metric is required to be free of co-ordinate
  singularities.
  \item $V_1+V_2<0$~: As previously shown, static orbifold solutions exist.  
 \end{enumerate}
\end{itemize}

\section{Bulk Slow--Roll for the Static Orbifold Solution}
\label{app:sr}

In this Appendix, we will solve the bulk equation~\eqref{eomf} for $\f$
for a static orbifold solution with $V>0$ and $V_1+V_2<0$. To do this,
we follow a procedure similar to the one explained in
section~\ref{slowroll}.  We start by introducing a new variable ${\cal
T}={\cal T}(\t ,t)$ with
\begin{equation}
 \frac{d\f}{dT} = -3\frac{\pt_\f V}{V}\; . \label{T1}
\end{equation}
We are now searching for solutions where the bulk scalar field depends
on $\t$ and $y$ only through this new variable, that is, $\f = \f
({\cal T}(\t ,y))$. It is straightforward to show from the $\f$
equation of motion~\eqref{eomf2} that, with the metric in conformal
gauge, ${\cal T}$ then satisfies the differential equation
\begin{equation}
 \ddot{\cal T}-T''+3\dot{\a}\dot{\cal T}-3\a '{\cal T}'=\frac{1}{3}e^{2\b}V
 \label{Tpde1}
\end{equation}
together with the boundary conditions
\begin{equation}
 \left[{\cal T}'\right]_{y=y_i} = \mp\left[e^\b\frac{\pt_\f V_i}{\pt_\f V}V
                          \right]_{y=y_i} \label{Tb1}
\end{equation}
provided that
\begin{equation}
 12\left(\frac{\pt_\f^2V\pt_\f V}{V^2}-\frac{(\pt_\f V)^3}{V^3}\right)
 (\dot{{\cal T}}^2-{{\cal T}'}^2)\ll e^{2\b}\pt_\f V\label{Tcond1}
\end{equation}
holds. This inequality will eventually lead to slow--roll conditions
once an explicit result for ${\cal T}$ is inserted. We observe that,
from eq.~\eqref{T1}, $\f$ should have a weak dependence on ${\cal T}$
for flat bulk potentials $V$. Finding an exact solution for ${\cal T}$
by solving eq.~\eqref{Tpde1} is quite difficult since the background
quantities $\a$ and $\b$ contain the various potentials which have a
time-variation through slow--roll of the scalar fields. However,
our strategy will be to solve eq.~\eqref{Tpde1} by assuming
that the potentials are effectively constant. This approximation
neglects higher-order contributions to the slow--roll evolution of $\f$
which, to leading order is governed by~\eqref{T1}. This, of course, means
that our final solution for $\f$ is only valid locally and describes
the $\t$ and $y$ variation of $\f$ around a given set of potential
values.

As we saw in section~\ref{slowroll}, even in this approximation, solutions
for ${\cal T}$ satisfying the boundary conditions~\eqref{Tb1} are
not easily found. However, for the $\s =0$ separating background~\eqref{met}
we can start with
\begin{equation}
 {\cal T} = \a +{\cal T}_0(\t )+{\cal T}_5(y)
\end{equation}
where the first term, $\a$, is a special inhomogeneous solution and the
other terms represent a separation Ansatz for the homogeneous part
of~\eqref{Tpde1}. The fact that ${\cal T}=\a$ is a special solution
of eq.~\eqref{Tpde1} can be shown rather generally by using the
background equations of motion~\eqref{eom00}-\eqref{eom05} and their first
integral~\eqref{int1}. The existence of this simple solution is of course
the prime motivation for introducing the variable ${\cal T}$.
For our specific separating background $\a$ takes the form
\begin{equation}
 \a = \a_0 +\x\t -\ln\cosh (\x y+\x_0)\; .
\end{equation}
The most general separating solution for the homogeneous part is given by
\bea
 {\cal T}_0 &=& C\x\t +Ae^{-3\x\t} \\
 {\cal T}_5 &=& BH(\x y+\x_0)+CI(\x y+\x_0)
\eea
where $A$, $B$, $C$ are arbitrary integration constants and the functions
$H$ and $I$ are defined by
\bea
 H(z) &=& (3+\sinh^2z)\sinh z \\
 I(z) &=& \ln\cosh z + \frac{1}{2}\cosh^2z +\sinh z\cosh^2z\;{\rm arctan}\,e^z
          +2\sinh z\;{\rm arctan}\,e^z\; .
\eea
We remark that $I$ is a special solution of
\begin{equation}
 \frac{d^2I}{dz^2}-3\tanh z\frac{dI}{dz} = 3
\end{equation}
while $H$ satisfies the homogeneous part of this equation.
We now have to implement the boundary conditions~\eqref{Tb1} which
fixes $B$ and $C$ to
\bea
 B &=& B_1\left(1-2\frac{\d_1}{\d}\right) +
       B_2 \left(1-2\frac{\d_2}{\d}\right) \\
 C &=& C_1\left(1-2\frac{\d_1}{\d}\right) +
       C_2 \left(1-2\frac{\d_2}{\d}\right)
\eea
where we recall that
\begin{equation}
 \d_i = \frac{\pt_\f V_i}{V_i}\; ,\qquad \d = \frac{\pt_\f V}{V}\; .
\end{equation}
Further $B_i$ and $C_i$ are complicated functions of $\r_i$ given by
\bea
 B_1 &=& D^{-1}\sqrt{1+\r_2^2}\, I'(-A_2)\r_1 \\
 B_2 &=& D^{-1}\sqrt{1+\r_1^2}\, I'(A_1)\r_2 \\
 C_1 &=& -3\, D^{-1}\left( 1+\r_2^2\right)^2\r_1 \\
 C_2 &=& -3\, D^{-1}\left( 1+\r_1^2\right)^2\r_2 \\
 D &=& 3\sqrt{1+\r_1^2}\sqrt{1+\r_2^2}\left[ \left( 1+\r_1^2\right)^{3/2}\,
       I'(-A_2) -\left( 1+\r_2^2\right)^{3/2}\, I'(A_1)\right]\; .
\eea
This result can now be used to derive the various slow--roll
conditions for the background solution under consideration by
inserting ${\cal T}$ into the inequalities~\eqref{bksr1}-\eqref{bksr3}
and \eqref{Tcond1}. Further conditions arise by inserting into the
equation of motion~\eqref{Tpde1} for ${\cal T}$ and by requiring that
additional terms proportional to time derivatives of potentials are
indeed small, as we have assumed. Clearly, this will lead to very
complicated slow--roll conditions, in general.  The above solution
significantly simplifies if expanded to lowest order in $\r_i$. Then
we arrive at the $\f$ evolution equation~\eqref{bsrsig0}  and the
slow--roll conditions~\eqref{bkeps1}-\eqref{bketa} which we have used
to analyze the specific examples in section~\ref{phen}.

\newpage




\begin{thebibliography}{99}

\bibitem{Horava:1996qa}
P.~Ho\v{r}ava and E.~Witten,
``Heterotic and type I string dynamics from eleven dimensions,''
Nucl.\ Phys.\ B {\bf 460} (1996) 506, hep-th/9510209.

\bibitem{Horava:1996ma}
P.~Ho\v{r}ava and E.~Witten,
``Eleven--dimensional supergravity on a manifold with boundary,''
Nucl.\ Phys.\ B {\bf 475} (1996) 94, hep-th/9603142.

\bibitem{Witten:1996mz}
E.~Witten,
``Strong coupling expansion of Calabi--Yau compactification,''
Nucl.\ Phys.\ B {\bf 471} (1996) 135, hep-th/9602070.

\bibitem{Lukas:1999yy}
A.~Lukas, B.~A.~Ovrut, K.~S.~Stelle and D.~Waldram,
``The universe as a domain wall,''
Phys.\ Rev.\ D {\bf 59} (1999) 086001, hep-th/9803235.

\bibitem{Lukas:1999tt}
A.~Lukas, B.~A.~Ovrut, K.~S.~Stelle and D.~Waldram,
``Heterotic M-theory in five dimensions,''
Nucl.\ Phys.\ B {\bf 552} (1999) 246, hep-th/9806051.

\bibitem{Lukas:1999hk}
A.~Lukas, B.~A.~Ovrut and D.~Waldram,
``Non--standard embedding and five--branes in heterotic M--theory,''
Phys.\ Rev.\ D {\bf 59} (1999) 106005, hep-th/9808101.

\bibitem{binf}
A.~Lukas, B.~A.~Ovrut and D.~Waldram,
``Boundary inflation,''
Phys.\ Rev.\ D {\bf 61} (2000) 023506, hep-th/9902071.

\bibitem{Nihei:1999}
T.~Nihei
``Inflation in the five--dimensional universe with an orbifold extra dimension,''
Phys.\ Lett.\ B {\bf 465} (1999) 81, hep-ph/9905487.

\bibitem{KK} 
H.~B.~Kim and H.~D.~Kim,
``Inflation and gauge hierarchy in Randall--Sundrum compactification,''
Phys.\ Rev.\ D {\bf 61} (2000) 064003, hep-th/9909053.

\bibitem{Binetruy:1999}
P.~Bin\'{e}truy, C.~Deffayet and D.~Langlois,
``Nonconventional cosmology from a brane universe,''
Nucl.\ Phys.\ B {\bf 565} (2000) 269, hep-th/9905012.

\bibitem{Ida}
D.~Ida,
``Brane world cosmology,''
JHEP {\bf 0009} (2000) 014, gr-qc/9912002.

\bibitem{Kanti:2000}
P.~Kanti, K.~Olive and M.~Popselov,
``Static Solutions for brane models with a bulk scalar field,''
Phys.\ Lett.\ B {\bf 481} (2000) 386, hep-ph/0002229.

\bibitem{Mohapatra:2000}
R.~N.~Mohapatra, A.~Perez--Lorenzana and C.~A.~de~Sousa~Pires,
``Cosmology of brane--bulk models in five--dimensions,''
Int.\ J.\ Mod.\ Phys.\ A {\bf 16} (2001) 1431, hep-ph/0003328.

\bibitem{Kim:2000}
J.~E.~Kim and B.~Kyae,
``Exact cosmological solution and modulus stabilization in the Randall--Sundrum model with bulk matter,''
Phys.\ Lett.\ B {\bf 486} (2000) 165, hep-th/0005139.

\bibitem{KSW}
J.~Khoury, P.~Steinhardt and D.~Waldram, 
``Inflationary solutions in the brane--world and their geometric interpretation,''
Phys.\ Rev.\ D {\bf 63} (2001) 103505, hep-th/0006069.

\bibitem{BCG} 
P.~Bowcock, C.~Charmousis and R.~Gregory, 
``General brane cosmologies and their global spacetime structure,''
Class.\ Quant.\ Grav. {\bf 17} (2000) 4745, hep-th/0007177.

\bibitem{Ochiai:2000}
H.~Ochiai and K.~Sato,
``Vacuum brane and the bulk dynamics in dilatonic brane world,''
Phys.\ Lett.\ B {\bf 503} (2001) 404, hep-th/0010163.

\bibitem{RS1}
L.~Randall and R.~Sundrum, 
``A large mass hierarchy from a small extra dimension,''
Phys.\ Rev.\ Lett. {\bf 83} (1999) 3370, hep-ph/9905221. 

\bibitem{MWBH}
R.~Maartens, D.~Wands, B.~Bassett and I.~Heard,
``Chaotic inflation on the brane,''
Phys.\ Rev.\ D {\bf 62} (2000) 041301, hep-ph/9912464.

\bibitem{CLL}
E.~Copeland, A.~Liddle and J.~Lidsey,
``Steep inflation: Ending brane world inflation by gravitational particle production,''
Phys.\ Rev.\ D {\bf 64} (2000) 023509, astro-ph/0006421.

\bibitem{nucleo1}
S.~Sarkar,
``Big bang nucleosynthesis and physics beyond the Standard Model,''
Rept.\ Prog.\ Phys. {\bf 59} (1996) 1493, hep-ph/9602260.

\bibitem{nucleo2}
K.~Olive, G.~Steigman and T.~Walker,
``Primordial nucleosynthesis: Theory and observations,''
Phys.\ Rep. {\bf 333} (2000) 389, astro-ph/9905320.  

\bibitem{Mukohyama:2000ui}
S.~Mukohyama,
``Gauge--invariant gravitational perturbations of maximally symmetric spacetimes,''
Phys.\ Rev.\ D {\bf 62} (2000) 084015, hep-th/0004067.

\bibitem{Kodama:2000fa}
H.~Kodama, A.~Ishibashi and O.~Seto,
``Brane world cosmology: Gauge--invariant formalism for perturbation,''
Phys.\ Rev.\ D {\bf 62} (2000) 064022, hep-th/0004160.

\bibitem{Langlois:2000ia}
D.~Langlois,
``Brane cosmological perturbations,''
Phys.\ Rev.\ D {\bf 62} (2000) 126012, hep-th/0005025.

\bibitem{vandeBruck:2000ju}
C.~van de Bruck, M.~Dorca, R.~H.~Brandenberger and A.~Lukas,
``Cosmological perturbations in brane--world theories: Formalism,''
Phys.\ Rev.\ D {\bf 62} (2000) 123515, hep-th/0005032.

\bibitem{Lyth}
D.~Lyth and A.~Riotto,
``Particle physics models of inflation and the cosmological density perturbation,''
Phys.\ Rep. {\bf 341} (1999) 1, hep-ph/9807278.

\bibitem{Lukas:1999kt}
A.~Lukas, B.~A.~Ovrut and D.~Waldram,
``Five--branes and supersymmetry breaking in M--theory,''
JHEP {\bf 9904} (1999) 009, hep-th/9901017.

\bibitem{Harvey:1999as}
J.~A.~Harvey and G.~Moore,
``Superpotentials and membrane instantons,''
hep-th/9907026.

\bibitem{Lima:2001jc}
E.~Lima, B.~A.~Ovrut, J.~Park and R.~Reinbacher,
``Non--perturbative superpotential from membrane instantons in heterotic  M-theory,''
hep-th/0101049.

\bibitem{Lukas:1998fg}
A.~Lukas, B.~A.~Ovrut and D.~Waldram,
``On the four--dimensional effective action of strongly coupled heterotic string theory,''
Nucl.\ Phys.\ B {\bf 532} (1998) 43, hep-th/9710208.

\bibitem{Lukas:1999ew}
A.~Lukas, B.~A.~Ovrut and D.~Waldram,
``The ten--dimensional effective action of strongly coupled heterotic  string theory,''
Nucl.\ Phys.\ B {\bf 540} (1999) 230, hep-th/9801087.

\bibitem{Bin}
P.~Bin\'{e}truy, C.~Deffayet, U.~Ellwanger and D.~Langlois,
``Brane Cosmological Evolution in a Bulk with a Cosmological Constant,''
Phys.\ Lett. {\bf B477} (2000) 285, hep-th/9910219.

\bibitem{pBB1}
M.~Gasperini and G.~Veneziano,
``Pre--big bang in string cosmology,''
Astropart.\ Phys. {\bf 1} (1993) 317, hep-th/9211021.

\bibitem{pBB2}
M.~Cavaglia,
``Pre big bang in M--theory,''
Class.\ Quant.\ Grav. {\bf 18} (2001) 1355, hep-th/0103095.

\bibitem{MaxiBoom}
The MaxiBoom Collaboration,
``CMB analysis of Boomerang \& Maxima \& the cosmic parameters $\Omega_{\rm tot},\Omega_{\rm b} h^2,\Omega_{\rm cdm} h^2,\Omega_{\Lambda},n_s$,''
Proc.\ IAU Symposium {\bf 201} PASP (2000) 65, astro-ph/0011378. 

\bibitem{reh1}
L.~Kofman, A.~Linde and A.~Starobinsky,
``Towards the theory of reheating after inflation,''
Phys.\ Rev.\ D {\bf 56}, (1997) 3258, hep-ph/9704452.

\bibitem{reh2}
Y.~Shtanov, J.~Traschen and R.~Brandenberger,
``Universe reheating after inflation,''
Phys.\ Rev.\ D {\bf 51} (1995) 5438, hep-ph/9407247.

\bibitem{TMM}
S.~Tsujikawa, K.~Maeda and S.~Mizuno,
``Brane preheating,''
Phys.\ Rev.\ D {\bf 63} (2001) 123511, hep-ph/0012141.

\bibitem{Dvali:1999}
G.~Dvali and G.~Gabadadze,
``Non--conservation of global charges in the brane universe and baryogenesis,''
Phys.\ Lett.\ B {\bf 460} (1999), 47, hep-ph/9904221.

\bibitem{CR}
H.~Chamblin and H.~Reall,
``Dynamic dilatonic domain walls,''
Nucl.\ Phys.\ B {\bf 562} (1999) 133, hep-th/9903225.

\bibitem{EKR}
K.~Enqvist, E.~Keski--Vakkuri and S.~Rasanen,
``Constraints on brane and bulk ideal fluids in Randall--Sundrum cosmologies,''
hep-th/0007254

\bibitem{SFV}
M.~Santos, F.~Vernizzi and P.~Ferreira,
``Isotropization and instability of the brane,''
hep-ph/0103112.


\end{thebibliography}
\end{document}